\documentclass[structabstract]{aa} % this is the printed version

\pdfoutput=1

\usepackage{natbib}
\bibpunct{(}{)}{;}{a}{}{,} % to follow the A&A style...

\usepackage{color}
\usepackage{amsmath,amssymb}
\usepackage{graphicx}
\usepackage[draft,linktocpage=true,unicode=true]{hyperref}
\usepackage{hypcap}

\newcommand{\ie}{\textit{i.e.}}

\newcommand{\lambdad}{{\ifmmode \lambda_{\rm D} \else $\lambda_{\rm D}$\fi}}
\newcommand{\pbar}{{\ifmmode \overline{p} \else $\overline{p}$\fi}}
\newcommand{\dbar}{{\ifmmode \overline{d} \else $\overline{d}$\fi}}
\newcommand{\gtilde}{{\ifmmode \tilde{\cal G} \else $\tilde{{\cal G}}$\fi}}
\newcommand{\gtildepos}{{\ifmmode \gtilde^{\; e^+} \else 
    $\gtilde^{\; e^+}$\fi}}
\newcommand{\gtildepbar}{{\ifmmode \gtilde^{\; \pbar} \else 
    $\gtilde^{\; \pbar}$\fi}}

\newcommand{\Msol}{{\ifmmode M_{\odot} \else $M_{\odot}$\fi}}
\newcommand{\Rsol}{{\ifmmode R_{\odot} \else $R_{\odot}$\fi}}
\newcommand{\rhosol}{{\ifmmode \rho_{\odot} \else $\rho_{\odot}$\fi}}
\newcommand{\fsol}{{\ifmmode f_{\odot} \else $f_{\odot}$\fi}}
\newcommand{\gev}{{\ifmmode {\rm GeV} \else ${\rm GeV}$\fi}}

\newcommand{\lcdm}{{\ifmmode \Lambda{\rm CDM} \else $\Lambda{\rm CDM}$\fi}}
\newcommand{\lsim}{\mathrel{\mathop{\kern 0pt \rlap
  {\raise.2ex\hbox{$<$}}}
  \lower.9ex\hbox{\kern-.190em $\sim$}}}
\newcommand{\gsim}{\mathrel{\mathop{\kern 0pt \rlap
  {\raise.2ex\hbox{$>$}}}
  \lower.9ex\hbox{\kern-.190em $\sim$}}}

\graphicspath{{./figs/}}
\begin{document}

\title{Galactic secondary positron flux at the Earth}
\author{
T. Delahaye\inst{1,2} \and F. Donato\inst{2} \and N. Fornengo\inst{2} \and J. Lavalle\inst{2} \and R. Lineros\inst{2} \and P. Salati\inst{1} \and R. Taillet\inst{1}
} 

%\offprints{\tt \\
%  delahaye@lapp.in2p3.fr\\
%  donato@to.infn.it\\
%  fornengo@to.infn.it\\
%  lavalle@to.infn.it\\
%  lineros@to.infn.it\\
%  salati@lapp.in2p3.fr\\
%  taillet@lapp.in2p3.fr}

\institute{LAPTH, Universit\'e de Savoie, CNRS, BP. 110, F-74941 Annecy-le-Vieux Cedex, France.
\and 
Dipartimento di Fisica Teorica, Universit\`a di Torino \& INFN - Sezione di Torino, Via P. Giuria 1, 10122 Torino, Italy.}

\date{Received 10 October 2008 / Accepted 4 March 2009}% It is always \today, today,

%--------------------------------------------------------------
\abstract{
%Text of Context (optional)
Secondary positrons are produced by spallation of cosmic rays within the interstellar gas.
Measurements have been typically expressed in terms of the positron fraction, which exhibits an increase above 10 GeV. 
Many scenarios have been proposed to explain this feature, among them some additional primary positrons originating from dark matter annihilation in the Galaxy.}{
%Text of aims
The PAMELA satellite has provided high quality data that has enabled high accuracy statistical analyses to be made, showing that the increase in the positron fraction extends up to about 100 GeV. 
It is therefore of paramount importance to constrain theoretically the expected secondary positron flux to interpret the observations in an accurate way.}{
%Text of methods
We focus on calculating the secondary positron flux by using and comparing different up-to-date nuclear cross--sections and by considering an independent model of cosmic ray propagation. 
We carefully study the origins of the theoretical uncertainties in the positron flux.}{
%Text of results
We find the secondary positron flux to be reproduced well by the available observations, and to have theoretical uncertainties that we quantify to be as large as about one order of magnitude. 
We also discuss the positron fraction issue and find that our predictions may be consistent with the data taken before PAMELA. 
For PAMELA data, we find that an excess is probably present after considering uncertainties in the positron flux, although its amplitude depends strongly on the assumptions made in relation to the electron flux.
By fitting the current electron data, we show that when considering a soft electron spectrum, the amplitude of the excess might be far lower than usually claimed.}{
%Text of Conclusions (optional)
We provide fresh insights that may help to explain the positron data with or without new physical model ingredients.
PAMELA observations and the forthcoming AMS-02 mission will allow stronger constraints to be aplaced on the cosmic--ray transport parameters, and are likely to reduce drastically the theoretical uncertainties.}

%--------------------------------------------------------------

\titlerunning{}
\authorrunning{Delahaye et al.}  
\maketitle

\begin{flushleft}
  A\&A 501, 821-833 (2009). \\
  Preprint LAPTH-1272/08 and DFTT 25/2008
\end{flushleft}

%--------------------------------------------------------------
\section{Introduction}
\label{sec:intro}

Among the different particles observed in cosmic rays, positrons still raise unanswered questions. 
Cosmic positrons are created by spallation reactions of cosmic ray nuclei with interstellar matter and propagate in a diffusive mode, because of their interaction with the turbulent component of the Galactic magnetic field. The expected flux of positrons can be calculated from the observed cosmic ray nuclei fluxes, using the relevant nuclear physics and solving the diffusion equation.

The HEAT experiment \citep{1997ApJ...482L.191B,2004PhRvL..93x1102B} showed that the positron fraction (the ratio of the positron o the total electron--positron fluxes) possibly exhibits an unexpected bump in the 10 GeV region of the spectrum.
Although this bump could be due to some unknown systematic effect, the HEAT result has triggered many explanations.
For instance, \cite{1998ApJ...493..694M} suggested that an interstellar nucleon spectrum harder than that expected could explain the excess.
Many works also focused on the dark matter hypothesis, the bump being due to a primary contribution from the annihilation of dark matter particles.
The positron excess expected in this framework is very uncertain, because the nature of dark matter is unknown, and the propagation of positrons involves physical quantities that currently are also not precisely known.
The related astrophysical uncertainties were calculated and quantified in~\citet{2008PhRvD..77f3527D}, where it was shown that they may be significant, especially in the low energy part of the spectrum, a property common also to the antiproton \citep[e.g.][]{2004PhRvD..69f3501D} and the antideuteron \citep{2000PhRvD..62d3003D,2008PhRvD..78d3506D} signals. For positrons, sizeable fluxes from dark matter annihilation are typically possible if dark matter overdensities are locally present, which is usually coded into the so--called ``boost factor''. A detailed analysis of the admissible boost factors for positrons and antiprotons was performed by \cite{2008A&A...479..427L}, who showed that boost factors are typically confined to less than about a factor of 10--20. Computing the antimatter fluxes directly in the frame of a cosmological N-body simulation leads to the same conclusions~\citep{2008arXiv0808.0332L}.

The PAMELA experiment \citep{2007APh....27..296P}  has released its first results on the positron fraction for energies ranging from 1.5 GeV to 100 GeV and with a large statistics \citep{PAMELA08}. The positron fraction is observed to rise steadily for energies above 10 GeV, reinforcing the possibility that an excess is actually present. It is therefore timely and crucial to complete a novel analysis of the positron flux, including a robust estimation of the accuracy of the theoretical determination. The calculation of the uncertainties affecting the standard spallation--induced positron population is especially important when unexpected distortions are observed in experimental data, to address the issue in a more robust way. This and an analysis of the positron signal from dark matter annihilation and its astrophysical uncertainties \citep{2008PhRvD..77f3527D}, will set the proper basis for discussing in detail the experimental results.

The uncertainties in the positron flux have several origins. First, the cosmic ray nuclei measurements have their experimental uncertainties, which then affect the predictions of induced secondary fluxes, such as positrons.
Second, various modelings of the nuclear cross--sections involved in the positron production mechanism are available, and they are not in complete agreement with each other, implying a range of theoretical variation.
Third, the uncertainties in the propagation parameters involved in the diffusion equation were thoroughly studied \citep{2001ApJ...555..585M}. A detailed analysis of their impact on the secondary positron flux is therefore needed.

In this paper, we therefore study the secondary positron production and transport in the Galaxy, with emphasis on determining the various sources of uncertainties, namely:
(i) the nuclear cross--sections,
(ii) the effect induced by the primary injection spectra,
(iii) the local interstellar medium, and
(iv) the propagation modeling.
The paper is organized into three main parts.
The positron injection spectrum and its uncertainties are derived in
Sect.~\ref{sec:sec_inj_spectrum}. We use up--to--date nuclear cross--sections and show the differences with older parameterizations.
The Green functions associated with the positron propagation throughout the Milky Way are discussed in Sect.~\ref{sec:propagation}. Our slab model is mostly characterized by energy losses and diffusion caused by magnetic turbulences.
The secondary positron flux at the Earth is presented in
Sect.~\ref{sec:positron_flux_uncertainties} with a range of variation
that includes the effects discussed throughout the paper.
We also confirm that considering the cosmic ray proton and $\alpha$
retro--propagation as well as diffusive reacceleration and convection
has little effect on our results above a few GeV.
Our results for the positron flux and its uncertainties agree with all the available measurements by different experimental collaborations (we recall that PAMELA does not provide, at the moment, the positron flux, but only the positron fraction). 
The positron flux, by itself, does not exhibit unusual features and is in basic agreement with the results of \cite{1998ApJ...493..694M} and \cite{2008ApJ...682..400P}. This could imply that both the HEAT excess and the PAMELA rise observed in the positron fraction, originate from dark matter annihilation, but we argue in Sect.~\ref{sec:positron_fraction} that electrons might also play an important role. Depending on the data set used to constrain the electron spectrum, the estimate of the positron fraction can exhibit quite different behaviors, which can affect the interpretation of the positron fraction data. We show that our predictions can indeed be consistent with the existing data of the positron fraction for a soft electron spectrum, compatible with fits to experimental data. Hard electron spectra, instead, definitely point toward the presence of an excess. We finally insist in Sect.~\ref{sec:conclusion}, as a main result of our analysis, that the current uncertainties in the transport parameters translate into one order of magnitude uncertainties in the secondary positron spectrum.

%--------------------------------------------------------------
\section{Production of positrons by spallation}
\label{sec:sec_inj_spectrum}

Secondary positrons are created by spallation of cosmic ray nuclei (mainly protons and helium nuclei) on interstellar matter (mainly hydrogen and helium).
We compute $q_{e^{+}}(\mathbf{x},E_{e})$, the number of positrons of energy $E_{e}$ created per unit volume at position $\mathbf{x}$, per unit time and per GeV. The positron source term reads:
\begin{eqnarray}
\label{eq:source}
  q_{e^{+}}(\mathbf{x},E_{e}) &=& 4 \pi
  \sum_{\text{targ=H,He}} \sum_{\text{proj}=p,\alpha}
  n_\text{targ}(\mathbf{x}) \\
  &\times & \displaystyle \int \Phi_\text{proj} \left( \mathbf{x} , E_\text{proj} 
  \right) \times dE_\text{proj} \times
  \frac{d\sigma}{dE_{e}}(E_\text{proj} \to E_{e}) , \nonumber
\end{eqnarray}
where $\Phi_\text{proj}\left( \mathbf{x} , E_\text{proj} \right)$ denotes the cosmic ray nucleon flux at position $\mathbf{x}$, $n_\text{targ}(\mathbf{x})$ the number density of target nuclei, and $d\sigma/dE_{e}$ the cross--section for the reactions creating positrons. We discuss these quantities this section.

%%-------------------------------------------------------------
\subsection{Spallation cross--sections for $p+p \to e^+$}

\begin{figure*}[t]
\begin{center}
\resizebox{\hsize}{!}{\includegraphics[width=\columnwidth]{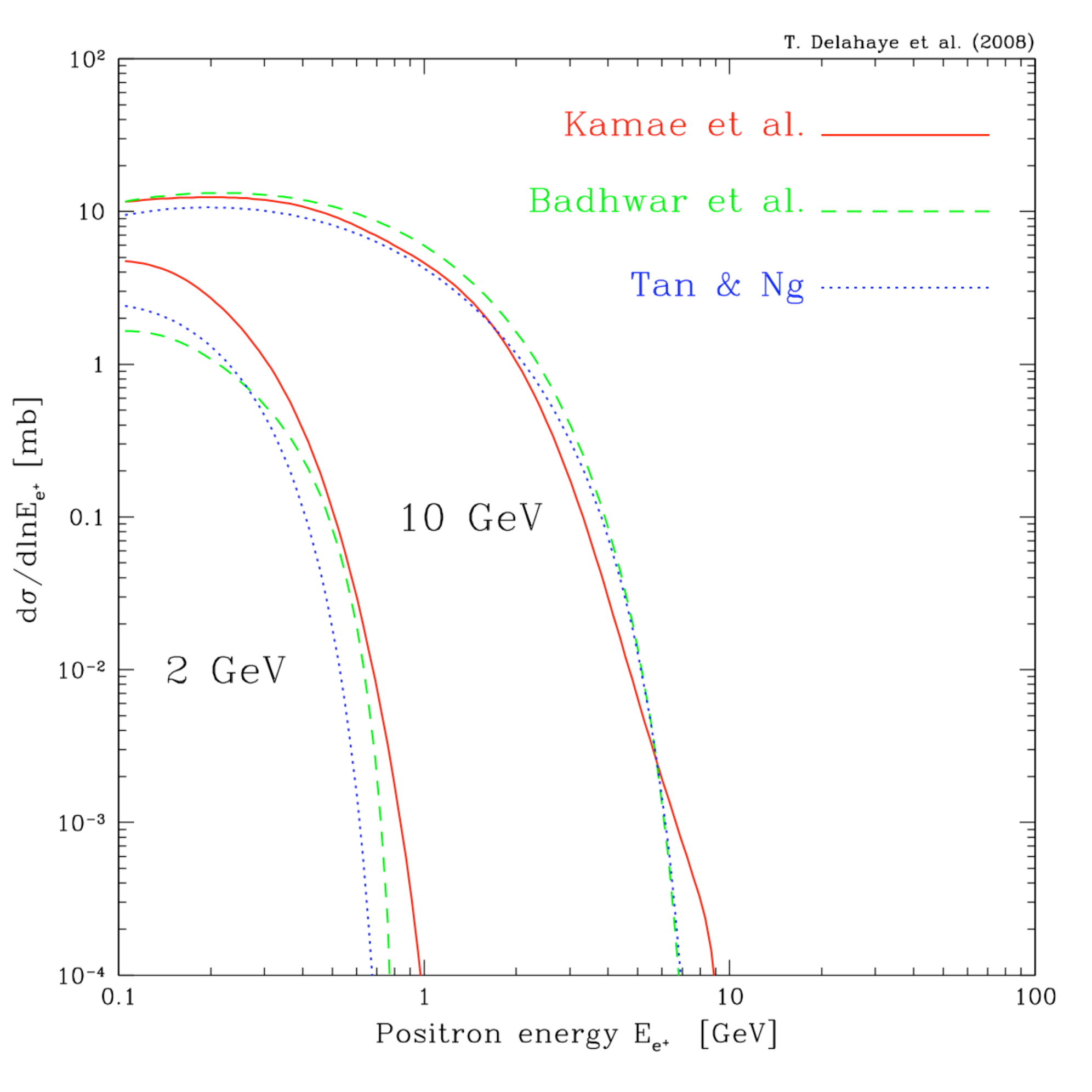}\includegraphics[width=\columnwidth]{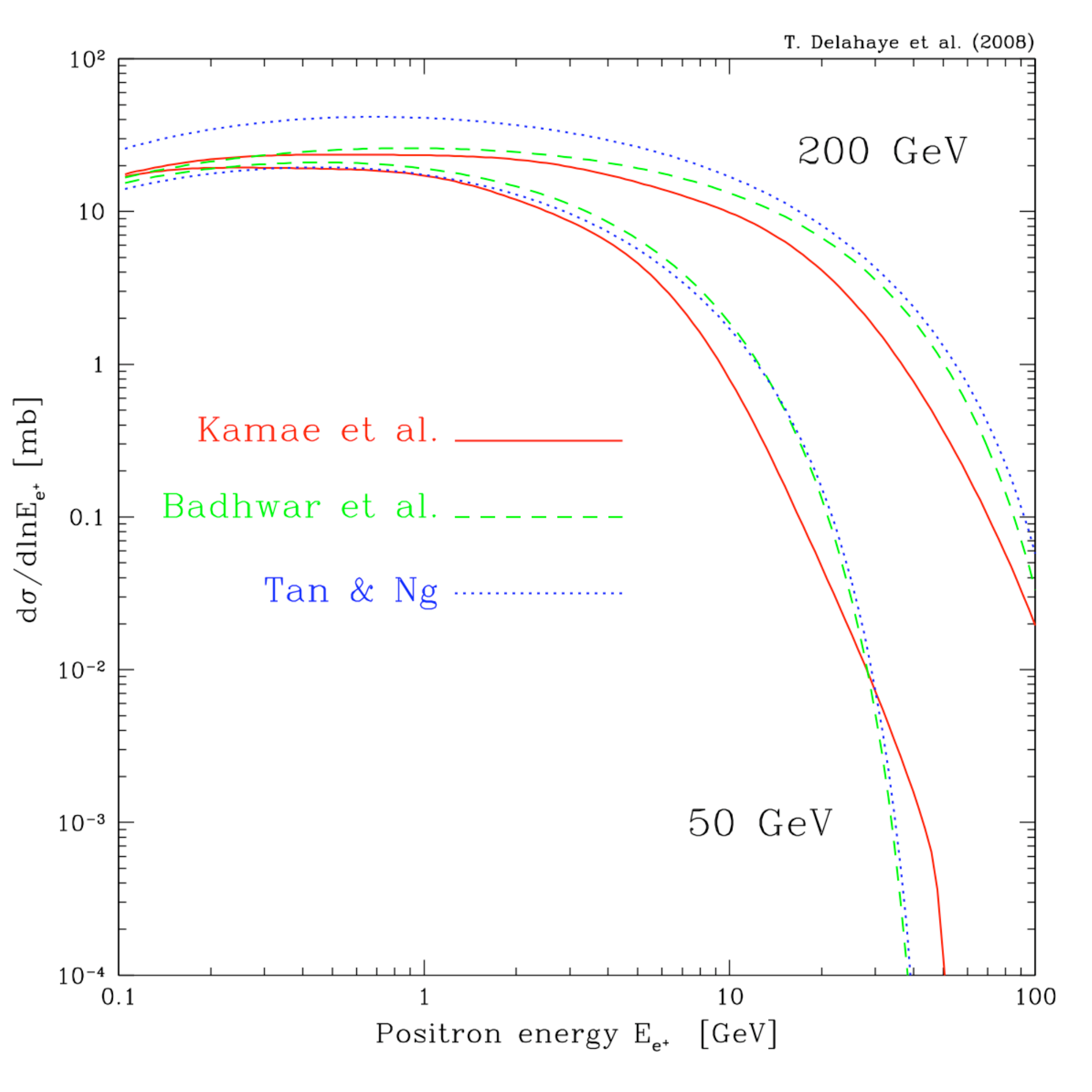}}
\caption{Comparison between various parameterizations of the positron production cross--section at different incident proton energies.} 
\label{fig:cross--section}
\end{center}
\end{figure*}

We first focus on the differential cross--section for the production of positrons. 
This production occurs by means of a nuclear reaction between two colliding nuclei, yielding mainly charged pions $\pi^\pm$ and other mesons, for which positrons are one of the final products of the decay chain. 
There are four main possible collisions: cosmic ray (CR) proton on interstellar (IS) hydrogen or helium; CR alpha particle on IS proton or helium. For the sake of clarity, we present only the formulae for the proton--proton collisions, but include all four processes in our results on the positron spectra.

At energies below about 3 GeV, the main channel for production of positrons involves the excitation of a Delta resonance, which then decays into pions:
\begin{equation}
  \text{p} + \text{H} \to \text{p} + \Delta^+
  \to \begin{cases} \text{p} + \pi^0 \\ \text{n} + \pi^+\end{cases} \;\; .
\end{equation}
The charged pions decay into muons, which subsequently decay into positrons. At higher energies direct production of charged pions proceeds with the process:
\begin{equation}
  \text{p} + \text{H} \to \text{p} + \text{n} + \pi^{+} \;\; .
\end{equation}
Kaons may also be produced:
\begin{equation}
  \textrm{p} + \textrm{H} \to X + K^\pm \;\; ,
\end{equation}
and the decay of kaons produces muons (63.44 \%) and pions (20.92 \%), which then decay into positrons as final products of their decay chain.

To compute the differential cross--section for the pion--production processes, we need the probability $d\sigma(E_p \to E_\pi)/dE_\pi$ of a spallation of a proton of energy $E_p$ yielding a pion with energy $E_\pi$ and the probability $\mathcal{P}(E_\pi \to E_{e})$ of such a pion eventually decaying into a positron of energy $E_{e}$. This second quantity can be computed thanks to basic quantum electrodynamics. whereas several parameterizations of the first quantity can be found in \cite{Badhwar1977}, \cite{tan_ng_1983_b} and \cite{Kamae2006}. The production cross--section of positrons is then given by:
\begin{eqnarray}
\frac{d\sigma}{dE_{e}}(E_\text{p} \to \pi^+ \to E_e) &=&
\nonumber \\
\int \frac{d \sigma}{d E_{\pi}}(E_\text{p} \to E_{\pi}) 
&\times& dE_{\pi} \times \mathcal{P}(E_{\pi} \to E_{e}) \; .
\label{pH_to_pi_to_e}
\end{eqnarray}
\cite{Kamae2006} also provided a direct parameterization of the 
$p + p \to e^+$ reaction
\footnote{
A few typos in the published version have been corrected
with the kind help of the authors.}.
All the afore mentioned parameterizations differ from each other and have been calibrated with different nuclear data sets. The default choice in our calculations, unless stated otherwise, is the \cite{Kamae2006} parameterization, which includes additional processes (especially resonances other than the Delta at low interaction energies) and has been calibrated with recent data. Ss stated in the original paper, we nevertheless warn that this parameterization relies on fits to Monte Carlo simulations and may be easilly affected by small uncertainties.

The cross--section for the process involving kaons can be computed in a similar way to the calculation of direct pion production. 
The QED expressions for the production of positrons from the kaon are provided e.g. in Appendix~D of \cite{1998ApJ...493..694M}.

In Fig. \ref{fig:cross--section}, we plot the cross--section for the positron production from the p-H scattering, as a function of positron energy. %
The incident proton energy is set to 2, 10 (left), 50 and 200 (right) GeV.
The three different plots at fixed proton energy correspond to the cross--section parameterizations of \cite{Kamae2006} (solid), \cite{Badhwar1977} (dashed) and \cite{tan_ng_1983_b} (dotted). The differences between these plots vary with both incident proton and final positron energies. For protons of intermediate energies of 10 and 50 GeV, the flatter part of the cross--section (at low energies) varies only slightly between the different parameterizations. The \cite{Kamae2006} parameterization undershoots and then overshoots the other two models only in the high energy range, while for slow protons (see the 2 GeV case), it provides more positrons. Because both the multiple baryonic resonances around 1600 MeV and the standard $\Delta$(1232) state have been added into their model. 

%%-------------------------------------------------------------
\subsection{Incident proton flux}
\label{subsec:nuclei}

\begin{figure}[t]
\begin{center}
\resizebox{\hsize}{!}{\includegraphics[width=\columnwidth]{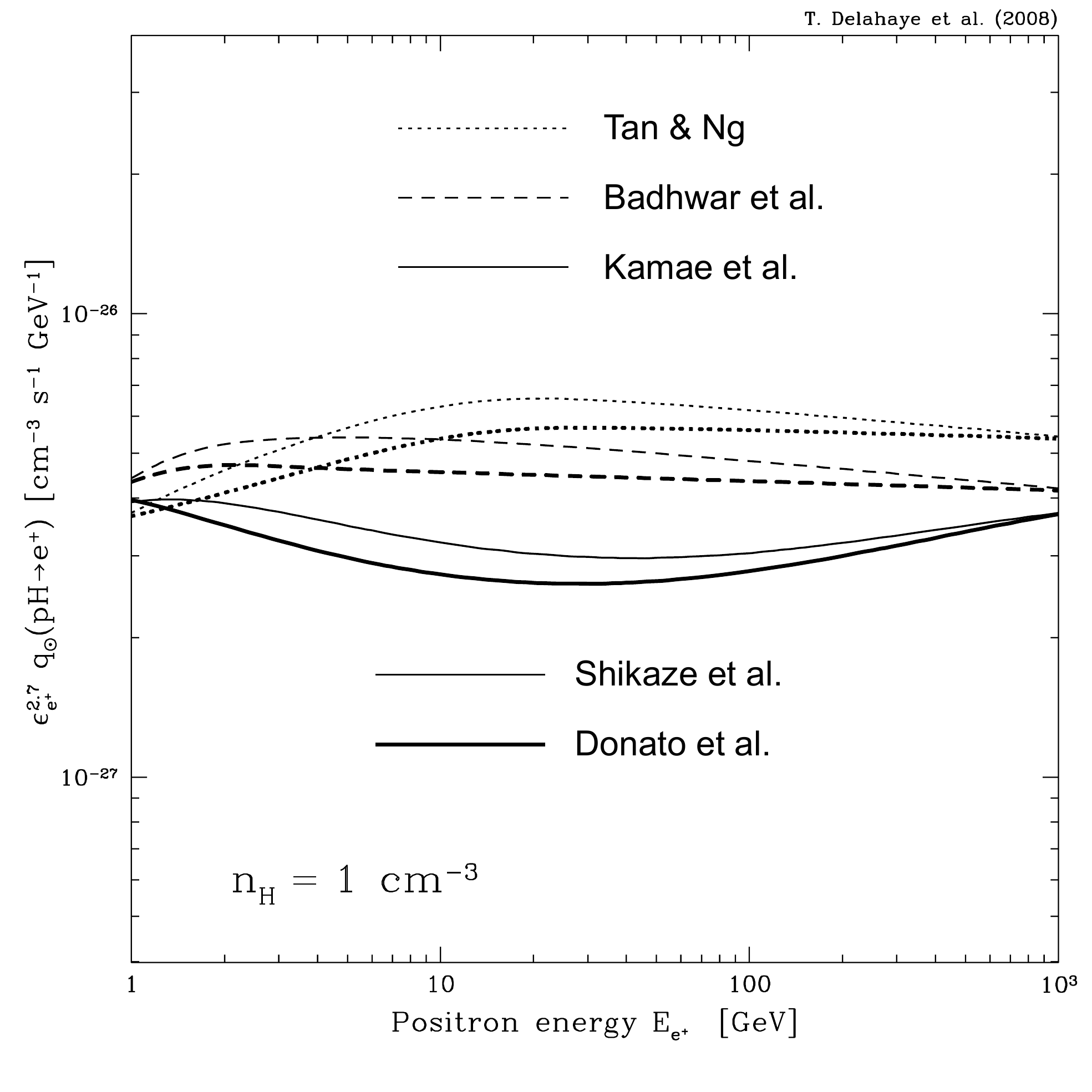}}
\caption{Comparison of the effect due to different parameterizations for the 
  cosmic ray proton spectra on the positron source term,
  as a function of the positrons energy. The additional effect induced
  by the different nuclear physics parameterizations is also shown. The 
  galactic protons density is taken at 1 hydrogen atom per cm$^{3}$.}
\label{fig:p_and_alpha_spectra}
\end{center}
\end{figure}

The proton flux $\Phi_\textrm{p}\left( \mathbf{x} , E_\textrm{p} \right)$ has been measured at the location of the Earth $\mathbf{x}=\mathbf{x}_\odot$ by several experiments. Various parameterizations of these measurements are found in the literature. In this analysis, we adopt the determinations of \cite{bess_shikaze_etal_07} and \cite{2001ApJ...563..172D}. The effect induced on the positron source term in Eq. (\ref{eq:source}) is  displayed in Fig.~\ref{fig:p_and_alpha_spectra}, where we also show the effect arising from the different nuclear physics parameterizations discussed in the previous subsection. The solid lines refer to \citet{Kamae2006}, the dashed lines to \citet{Badhwar1977}, and the dotted lines to \citet{tan_ng_1983_b}. For each set of curves, the thick lines are obtained for the proton spectrum parameterization of \citet{2001ApJ...563..172D}, while the thin lines refer to \cite{bess_shikaze_etal_07}. Figure \ref{fig:p_and_alpha_spectra} illustrates that different parameterizations of the incident proton flux cause less significant uncertainty than the nuclear models. At low (1 GeV) and high (10$^{3}$ GeV) energies, the \cite{bess_shikaze_etal_07} and \cite{2001ApJ...563..172D} fluxes produce almost identical results on the proton flux, but at intermediate energies the differences are also negligible.

In contrast, the cross--section parameterizations produce more significant variations. Figure \ref{fig:p_and_alpha_spectra} shows that the model of \citet{Kamae2006} gives the faintest spectrum in all the energy range (from 1 GeV to 1 TeV), while that of \citet{tan_ng_1983_b} corresponds to the maximal spectrum. The most significant difference occurs at 20--30 GeV, of about a factor of two.

The spatial dependence of the proton flux is determined by solving the diffusion equation and normalizing the resulting flux to the solar value. This procedure will be referred to as retro--propagation in the following. We find that retropropagation is not crucial and that one can safely approximate the proton flux as being homogeneous and equal to the solar value. This is because most positrons detected in the solar neighborhood have been created locally, over a region where the proton flux does not vary significally. This will be discussed further in the final section.

For the sake of simplicity, we show only the impact of the pH interaction on the determination of the positron spectrum, but we recall that our results in the subsequent sections refer to cosmic ray protons and $\alpha$ interacting with both IS hydrogen and helium with densities of $n_{\text{H}} = 0.9$ cm$^{-3}$ and $n_{\text{He}} = 0.1$ cm$^{-3}$ respectively. These average values for the hydrogen and helium densities are of course approximations based on local estimates, but we do not expect significant changes in the averaging within the kpc scale  \citep{2007A&A...467..611F}, which, as we show in the subsequent sections, is the relevant scale for the secondary positron propagation. Cross--sections for heavy nuclei were dealt with in the same way as in \cite{pi_nuclear_norbury_07}.

%--------------------------------------------------------------
\section{Propagation}
\label{sec:propagation}

In the Galaxy, a charged particle travelling between its source and the solar neighborhood is affected by several processes.
Scattering by magnetic fields leads to a random walk in both real space (diffusion) and momentum space (diffusive reacceleration).
Particles may also be spatially convected away by the galactic wind (which induces adiabatic losses), and lose energy as they interact with either interstellar matter or the electromagnetic field and radiation of the Galaxy (by synchrotron radiation and inverse Compton processes).
Above a few GeV, the propagation of positrons in the Milky Way is dominated by space diffusion and energy losses.\\

In this paper, the diffusion coefficient is assumed to be homogeneous and isotropic, with a dependence on energy given by
$K(E) = \beta \, K_0 \, (\mathcal{R}/1 \; {\rm GV})^{\delta}$,
where the magnetic rigidity 
$\mathcal{R}$ is related to the momentum $p$ and electric charge $Ze$
by $\mathcal{R}=pc/Ze$. 
Cosmic rays are confined within a cylindrical diffusive halo of radius $R = 20 \;\text{kpc}$ and height $2L$, their density vanishing at the boundaries $N(|z|=L,r)=N(z,r=R)=0$. 
As discussed below, the radial boundary has a negligible effect on the density of positrons in the Solar System.
In this section, we do not consider the possibility of Galactic convection and diffusive reacceleration. 
These processes are taken into account in Sect.~\ref{sec:diffusive_reacceleration}, where we demostrate that they have little effect.
The convective wind is assumed to carry cosmic rays away from the Milky Way disk in the $z$ direction at a constant velocity $V_c$. 
Diffusive reacceleration depends on the velocity $V_{a}$ of the Alfv\`en waves.
The free model parameters are therefore the size $L$ of the diffusive halo, both the normalization $K_0$ and spectral index $\delta$ of the diffusion coefficient, the convective wind velocity $V_{c}$, and the Alfv\`en velocity $V_{a}$ (see Sect.~\ref{subsec:propag_param} for additional details).
This model has been consistently used in several studies to constrain the propagation parameters
\citep{2001ApJ...555..585M,2002A&A...394.1039M,2002A&A...381..539D} and examine their consequences \citep{2003A&A...402..971T,2003A&A...404..949M} for the standard \pbar\ flux \citep{2001ApJ...563..172D}, the exotic \pbar\ and \dbar\ fluxes \citep{toolreview, 2006astro.ph..9522M,2004PhRvD..69f3501D,2002A&A...388..676B,2005PhRvD..72f3507B,2007PhRvD..75h3006B}, and also for positrons~\citep{2008A&A...479..427L,2008PhRvD..77f3527D}.
The reader is referred to \citet{2001ApJ...555..585M} for a more detailed presentation and motivation of the framework.

%%---------------------------------------------------------------------
\subsection{The Green function for positrons}
\label{subsec:green_function}

The propagation of positrons differs from that of the nuclei in several respects.
Although space diffusion is an essential ingredient common to all cosmic ray species, positrons suffer mostly from inverse Compton and synchrotron energy losses, e.g. \citealt{1998ApJ...493..694M}, whereas \mbox{(anti--)protons} are mostly sensitive to the galactic wind and the nuclear interactions as they cross the Milky Way disk.
As a result, a positron line at source leads to an extended spectrum after
its propagation. 
This disagrees with studies of most nuclear species for which, as first approximation, energy losses can be neglected.
Consequently, the diffusion equation that leads to the positron number density $N$ per unit of volume and energy, with the source term $q_{e^{+}}(\mathbf{x},E)$, becomes:
\begin{equation}
\label{eq:untilded}
 -K_0 \left(\frac{E}{E_0}\right)^\delta \triangle N  + 
 \frac{\partial}{\partial E} \left\{ \frac{dE}{dt}\;  N \right\} = 
 q_{e^{+}}(\mathbf{x},E).
\end{equation}
The first term is simply the diffusion coefficient written as $K({\cal R})\approx K_0 (E/E_0)^\delta$, where $E_0\equiv 1$ GeV is used 
to maintain the correct units throughout the paper.\\
The synchrotron and inverse Compton losses can be written as $dE/dt(E) = - E^{2}/(E_0 \tau_{E})$.
Defining a pseudo-time:
\begin{equation}
\hat{t} \equiv \tau_E \; \frac{{(E/E_0)}^{\delta-1}}{1-\delta}
\label{eq:t_hat}
\end{equation}
and applying the following rescaling:
\begin{equation*}
	\hat{N} \equiv {(E/E_0)}^{2} N \quad \text{and}\quad
	\hat{q}_{e^{+}}(\mathbf{x},E) \equiv {(E/E_0)}^{2-\delta} 
        q_{e^{+}}(\mathbf{x},E),
\end{equation*}
the diffusion equation can be rewritten as:
\begin{equation}
\label{eq:tilded}
\frac{\partial \hat{N}}{\partial \hat{t}} - K_0 \, \triangle \hat{N} =
\hat{q}_{e^{+}}(\mathbf{x},\hat{t}) \; ,
\end{equation}
which is formally identical to the well--known time-dependent diffusion equation \citep{bulanov74,1999PhRvD..59b3511B}.\\

It proves convenient to separate diffusion along the radial and vertical directions. 
Considering a source located at $(x,y,z,\hat{t}_{E_S})$ and detected at $(\Rsol,0,0,\hat{t}_{E_O})$, the corresponding flux depends only on the radial relative distance $r  = |{\bf r}_S-{\bf r}_O|$, the distance of the source from the plane $z=z_S$ and the relative pseudo--time $\hat{\tau}=\hat{t}_{E_O}-\hat{t}_{E_S}$. When the radial boundary is taken into account, one has to use an expansion over Bessel functions \citep{bulanov74,2008PhRvD..77f3527D}. However, in most situations it is safe to ignore the radial boundary. The Green function 
${\cal \hat{G}}_\odot(r ,z,\hat{\tau})$ of Eq.~(\ref{eq:tilded}) is then given 
by:
\begin{equation}
\label{eq:pseudo-propag}
{\cal \hat{G}}_\odot(\hat{\tau},r ,z)=
{\displaystyle
\frac{\theta \left( \hat{\tau} \right)}{ 4 \pi K_{0} \hat{\tau}}}
\; \exp \left(-\frac{r^{2}}{4 K_{0} \hat{\tau}} \right) \,\times \,
{\cal G}^{1D}(z,\hat{\tau}) \; .
\end{equation}
The radial behavior of the positron Green function encourages us to define
the characteristic diffusion length:
\begin{equation}
\lambdad \equiv \sqrt{4 K_0  \hat{\tau}}\, .
\label{eq:ld}
\end{equation}
This length defines the scale of the positron sphere, i.e, the region where most of the positrons detected at the Earth are produced.
It depends on the injected ${E_S}$ and detected ${E_O}$ positron energies in terms of the pseudo--time difference $\hat{\tau}$. 
For GeV energies, $\lambdad \lesssim 5$ kpc, typically, which defines the local 
character of the positron origin.

The effect of boundaries along $z=\pm L$ appears in only ${\cal G}^{1D}(z,\hat{\tau})$. For convergence properties, two distinct regimes are worth considering~\citep{2007A&A...462..827L}
\begin{enumerate}
\item When the extension $\lambdad$ of the positron sphere is smaller
  than the half--thickness $L$ of the diffusive halo, it is most appropriate to use
  the so-called electrical image formula (e.g.~\citealt{1999PhRvD..59b3511B}):
  \begin{equation}
      {\cal G}^{1D}(z,\hat{\tau})\!\!= \!\!\!
      { \sum_{n = -\infty}^{+\infty}} \!\! \left( -1 \right)^{n} 
      \frac{\theta \left( \hat{\tau} \right)}{\sqrt{ 4 \pi K_{0} 
	  \hat{\tau}}} \;
      \exp \left\{ - \frac{\left( z_n - z \right)^{2}}{4 K_{0} 
	\hat{\tau}}\right\},
      \label{eq:V_image}
   \end{equation}
   where $ z_n = 2 L n + \left( -1 \right)^{n} z$;

 \item In the opposite situation, a more suitable expression is based on an analogy with the solution to the Schr\"odinger equation in an infinitely deep square potential: expansion of the solution over the eigenfunctions of the Laplacian operator:
   \begin{eqnarray}
	 {\cal G}^{1D}(z,\hat{\tau})& = &
	 \frac{1}{L}\! \sum_{n=1}^{+ \infty} 
	 e^{- K_0 k_n^2 \hat{\tau}} \phi_{n}(0) \phi_{n}(z)
	  \nonumber\\
	 & & + e^{-  K_0 {k_n^\prime}^2 \hat{\tau}} \phi_n^\prime(0) 
          \phi_n^\prime(z)\!
	 \label{eq:V_quantum}
   \end{eqnarray}
   where:
   \begin{eqnarray}
     \phi_n(z) = \sin \left[ k_n (L-|z|) \right]  \;&\text{;}&\; k_n = 
     \left( n - \frac{1}{2} \right)\frac{\pi}{L}\;\; {\rm (even)} \nonumber\\
	 \phi_n^\prime(z) = \sin\left[ k_n^\prime (L-z)  \right]  &\text{and}&  
	 k_n^\prime = n \frac{\pi}{L} \;\; \text{(odd).} \nonumber
   \end{eqnarray}
\end{enumerate}

The true positron propagator for a monochromatic point source is related to Eq.~(\ref{eq:pseudo-propag}) by 
\begin{equation}
 {\cal G}^{\; e^+}_\odot(E\leftarrow E_S,r ,z)= \frac{\tau_E E_0}{E^2} \times 
 {\cal \hat{G}}_\odot(\hat{\tau}=\hat{t}_E-\hat{t}_{E_S},r ,z)\;.
\label{eq:propa}
\end{equation}
The secondary positron flux at the Earth is then given by (considering the Earth as the origin of the coordinate system):
\begin{eqnarray}
\phi_{e^{+}}^{\odot}(E) & = & \frac{\beta c}{4\pi} \times
\int_{E}^{\infty} dE_S \times \label{eq:secflux} \\
& \times &
\int_{\rm slab}d^3{\bf x}_S \,
{\cal G}^{\; e^+}_\odot(E\leftarrow E_S,r_S ,z_S) \,
{q}_{e^{+}}(\mathbf{x}_S,E_S) \; , \nonumber
% \frac{dn_{e^+}^{\rm source}}{dE_S}({\bf x}_S,E_S)\;,
\end{eqnarray}
where $|{\bf x}_S|^2 = r_S^2 + z_S^2$ is the squared distance from the source to the Earth, and $\beta$ is the positron velocity in units of the velocity of light.\\

%%-------------------------------------------------------------------
\subsection{Analytical solution for a homogeneous source term}
\label{subsec:analytical_sol}

The secondary positrons originate from spallation processes of cosmic rays off the interstellar gas, the latter being located mainly inside the galactic disk.
If we assume that the cosmic ray fluxes and the gas are homogeneous inside the disk and place the radial boundaries at infinity (see Sect.~\ref{subsec:green_function}), the spatial integral in Eq.~(\ref{eq:secflux}) can be derived analytically, e.g. by implementing an infinite disk of thickness $\lbrack -z_{\rm max},z_{\rm max}\rbrack$.
The flux at the Earth simplifies into
\begin{equation}
\phi_{e^{+}}^{\odot}(E) = \frac{\beta c}{4\pi}
\int_{E}^{\infty} dE_S \;
q_{e^{+}} \!\! \left( E_{S} \right) \times \frac{\tau_E E_0}{E^2}
\times \eta(\lambdad) \, ,
\label{eq:secondary_flux_simple}
\end{equation}
where
\begin{equation}
\eta(\lambdad) = \int_{-z_{\rm max}}^{+z_{\rm max}} d^3{\bf x}_S \;
{\cal \hat{G}}_\odot(\hat{\tau}=\hat{t}_E-\hat{t}_{E_S},r ,z)\;.
\label{eq:def_eta}
\end{equation}
This quantity depends only on the characteristic scale $\lambdad$ and can be expressed as
\begin{equation}
\eta(\lambdad) =
\begin{cases}
{\displaystyle
\frac{1}{2}\sum_{n=-\infty}^{\infty} \left\{ {\rm Erf} \left( 
  \frac{z_n^{\rm max}}{\lambdad}\right) -  
  {\rm Erf} \left( \frac{z_n^{\rm min}}{\lambdad}\right) \right\} \; ,}\\
{\displaystyle
  \frac{2}{L}\sum_{n=1}^{\infty}(-1)^{n+1}
  \frac{\cos \left( k_n(L-z_{\rm max}) \right) }{k_n}
  \times e^{-k_n^2\lambdad^2/4 } \; .}
\end{cases}
\label{eq:analytique_julien}
\end{equation}
The upper line corresponds to the electrical image solution, while the lower line is obtained by expanding the solution over the eigenfunctions of the Laplacian operator (see Sect.~\ref{subsec:green_function}).
Regarding the former case, we define
$z_{n}^\text{max}\equiv 2 n L + (-1)^n z_\text{max}$ and
$z_{n}^\text{min}\equiv 2 n L - (-1)^n z_\text{max}$ where
$z_\text{max}$ is defined to be 100~pc.
The energy dependence of the solutions is hidden in the propagation length \lambdad.
This simplification is helpful because it can provide an efficient way of checking the numerical spatial integral. 
Moreover, if the source term is shown to be almost homogeneous in the thin galactic disk, 
then this solution fully applies. 
Homogeneity is certainly justified in the energy range considered here but could break down at TeV energies.

\begin{figure}[t]
\begin{center}
\resizebox{\hsize}{!}{\includegraphics[width=\columnwidth]{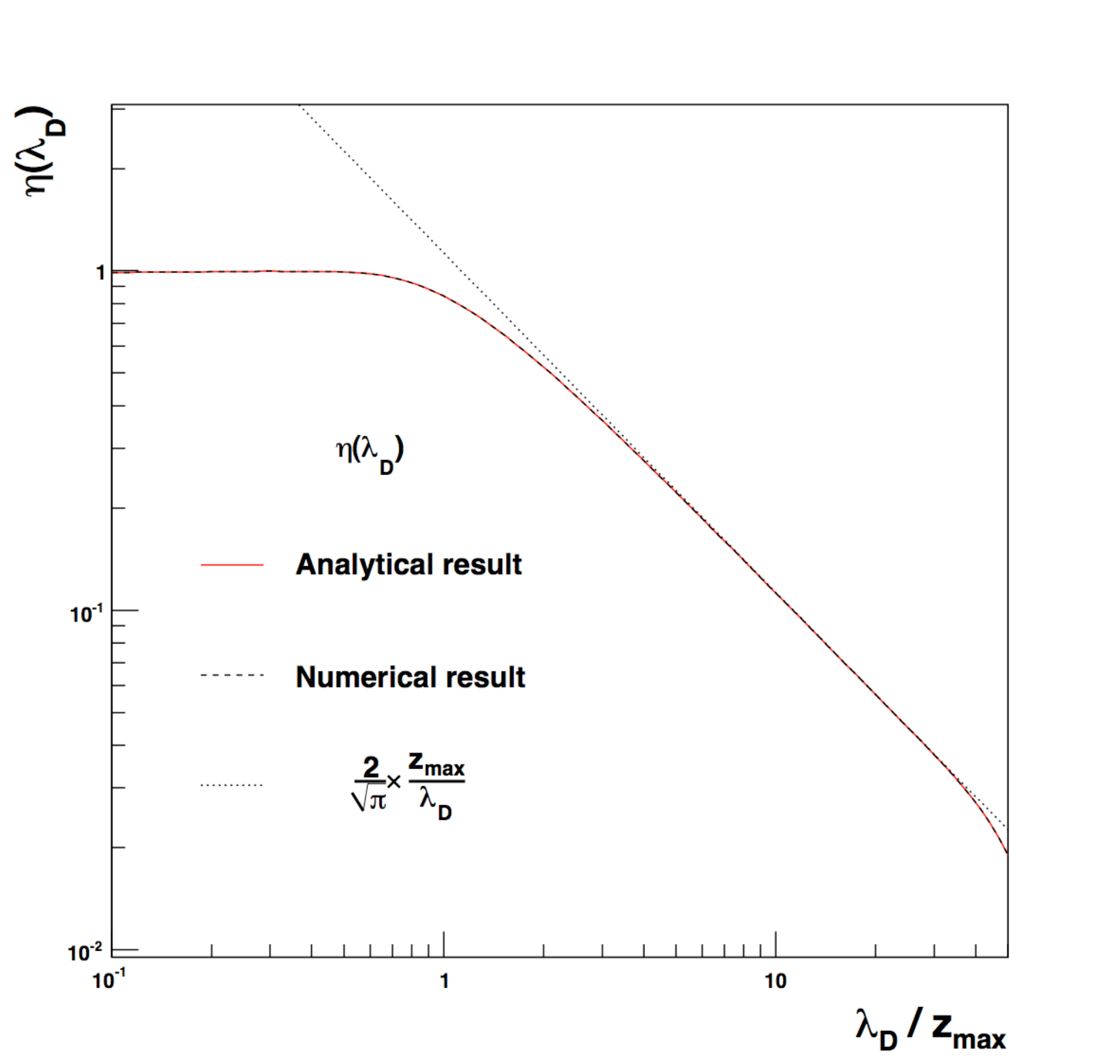}}
\caption{The integral $\eta$ is plotted as a function of the ratio
${\lambdad}/{z_\text{max}}$. Because $\eta$ can be interpreted as the
the fraction of the positron sphere intersected by the Galactic disk,
we infer that it should be unity for ${\lambdad} \ll {z_\text{max}}$.
In the converse regime, $\eta$ is proportional to the ratio
${z_\text{max}}/{\lambdad}$. See text for further details.} 
\label{fig:eta}
\end{center}
\end{figure}

In Fig.~\ref{fig:eta}, $\eta$ is plotted as a function of the propagation length $\lambdad$. 
This parameter can be interpreted as the ratio of received to produced positrons. 
In the regime where $\lambdad$ is much smaller than the half--thickness $L$ of the diffusive halo, positron propagation occurs as if the diffusive halo were infinite. 
In this 3D limit, the propagator ${\cal \hat{G}}_\odot(\hat{\tau}=\hat{t}_E-\hat{t}_{E_S},r ,z)$
describes the probability of a positron detected at Earth originating in the location $r$ and $z$ and as such is normalized to unity.
Secondary positrons are produced solely in the disk and $\eta$ equals by definition the fraction of the positron sphere filled by the disk.
We therefore expect this fraction to be close to 1 when $\lambdad$ is smaller than $z_\text{max}$ as the positron sphere becomes small enough to be embedded entirely inside the Galactic disk. 
In the converse situation, $\eta$ decreases approximately as
\begin{equation}
\eta ( \lambdad \gg z_\text{max} ) \simeq
\frac{2}{\sqrt{\pi}} \, \frac{z_\text{max}}{\lambdad} \; .
\end{equation}
Both behaviors are illustrated in Fig.~\ref{fig:eta}.

%%-------------------------------------------------------------------
\subsection{Propagation parameters}
\label{subsec:propag_param}

The propagation parameters $\delta$, $K_0$, $L$, $V_c$, and $V_{a}$ are not measured directly. 
However, it is possible to determine the parameter sets that are consistent with the observed properties of nuclei cosmic ray spectra, by comparing the observed spectra with the predictions of the diffusion model.
In \citet{2001ApJ...555..585M,2002A&A...394.1039M}, the secondary--to--primary ratio B/C is used to place constraints on the parameter space, and in \citet{2002A&A...381..539D} the information provided by the cosmic ray radioactive species is studied.
For the discussion below, it is convenient to isolate three sets of parameters labeled MIN, MED, and MAX, defined in Table~\ref{table:prop}.
These configurations are named according to the primary antiproton signal yielded by dark matter species annihilating in the Milky Way halo \citep{2002A&A...388..676B,2004PhRvD..69f3501D} and filling completely the diffusive halo.
The half--thickness $L$ increases from 1 to 15~kpc between models MIN and MAX.
Notice that in the B/C analysis of \citet{2001ApJ...555..585M}, $L$ is shown
to be correlated with the normalization $K_{0}$. 
Thick diffusive halos are associated with high diffusion coefficients and hence to high values of $\lambdad$. 
Inspired by Fig.~\ref{fig:eta}, we anticipate that the
secondary positron flux should have its lowest value in the MAX configuration
and its highest for the MIN model.
%
%TTTTTTTTTTTTTTTTTTTTTTTTTTTTTTTTTTTTTTTTTTTTTTTTTTTTTTTTTTTTTTTTTTTTTTTTTTTTTTTTTTTTTTTTTT
\begin{table}[t]
\begin{center}
\begin{tabular}{|c||c|c|c|c|c|c|}
\hline
Model  & $\delta$ & $K_0$ & $L$ & $V_{c}$ & $V_{a}$ \\
       &          & [kpc$^2$/Myr] & [kpc] & [km/s] & [km/s] \\
\hline
MIN  & 0.85 &  0.0016 & 1  & 13.5 &  22.4 \\
MED  & 0.70 &  0.0112 & 4  & 12   &  52.9 \\
MAX  & 0.46 &  0.0765 & 15 &  5   & 117.6 \\
\hline
\end{tabular}
\vskip 0.25cm
\caption{
Typical combinations of diffusion parameters that are compatible with the B/C
analysis \citep{2001ApJ...555..585M}. As shown in \cite{2004PhRvD..69f3501D}, 
these propagation models correspond respectively to minimal, medium, and maximal 
primary antiproton fluxes.}
\label{table:prop}
\end{center}
\end{table}

\subsection{Energy losses}

A detailed analysis including all energy--loss processes is presented in Sect.~\ref{sec:diffusive_reacceleration}. Here, we focus on the main processes for our energy range of interest. 
At energies higher than about 10 GeV, the most relevant energy losses are those caused by synchrotron radiation and inverse Compton (IC) scattering of the cosmic 
microwave background (CMB) and stars photons:
\begin{equation}
-b^{\rm loss}(\epsilon) = \frac{\epsilon^2}{\tau_E} = 
\frac{\epsilon^2}{\tau_{\rm sync}} + \frac{\epsilon^2}{\tau_\ast} + 
\frac{\epsilon^2}{\tau_{\rm CMB}}. 
\label{eq:b_loss}
\end{equation}
Each energy--loss timescale $\tau$ can be calculated by means of both the 
Compton cross--section and the corresponding radiation--field energy density, 
as detailed in detail by ~\cite{1994hea..book.....L}.
These Compton processes can indeed be computed in the Thomson limit. 
In the energy domain of interest here for positrons, say 10-100 GeV, the condition $\gamma_{e^+} E_{\rm ph} < m_e$, where $\gamma_{e^+}$ is the positron Lorentz factor and $E_{\rm ph}$ is the photon energy, is fulfilled for all relevant frequencies of the considered radiation fields.

By studying Eqs.~(\ref{eq:t_hat}) and  (\ref{eq:ld}), we note that $\tau_E$ has a significant effect on the positron propagation length: the higher the value of $\tau_E$, the farther is the origin of detected positrons from Earth.
Since the Green function has a Gaussian dependence on the propagation scale, the importance of estimating accurately the $\tau_E$ parameter is reinforced, even though we expect the effect to be lower for secondaries than for primaries, for which the source term may, in some cases, exhibit a far more pronounced spatial dependency. 
In the following, we briefly discuss the different contributions to the energy--loss timescale in the high energy range.
The CMB contribution is the most straightforward to compute. Adopting a mean CMB temperature of $T = 2.725$~K, we readily determine $\tau_{\textnormal{CMB}} = 3.77~\times~10^{16}$s.\\

For the synchrotron contribution, the local value of the magnetic field, which equals $\tau_{\textnormal{sync}}$, remains unknown.
As explained in \cite{2003A&A...411...99B}, there are two different methods of measurement.
The first relies on the intensity of the synchrotron radiation from cosmic electrons and gives a value of $B \sim 4 \pm 1 \; \mu$~G (\cite{Prouza:2003yf}). % ** 
However, this value depends on the adopted model, particularly in terms of the cosmic ray electron spectrum estimation.
The second method uses the Faraday rotation measurements of pulsar polarized emission, and yields $B \sim 1.8 \pm 0.3 \; \mu$~G (\cite{2006ApJ...642..868H}).
The two results are inconsistent but \cite{2003A&A...411...99B} were able to identify further uncertainty in the second method, which, if the thermal electron density is anticorrelated with the magnetic field strength, produces a revised value of $B \in [1.5 ; 4 ] \; \mu$~G, and hence $\tau_{\textnormal{sync}} \in [2.47 \times 10^{16} ; 1.76 \times 10^{17}]$~s.

Finally, to evaluate the contribution of IC processes in the interstellar radiation field (ISRF), we rely on the study performed by \cite{2000ApJ...537..763S}.
They estimated the local value of the ISRF energy density, whose uncertainty can be 
inferred from its variation in value within the 2 kpc around the solar position. 
This provides an average local ISRF energy density of $U_{\textnormal{rad}} \sim 2 \pm 1$~eV~cm$^{-3}$, hence $\tau_{\textnormal{IC}} \in [6.54 \times 10^{15} ; 1.96 \times 10^{16}]$~s.

\begin{figure}[t]
\begin{center}
\resizebox{\hsize}{!}{\includegraphics[angle=90,width=\columnwidth]{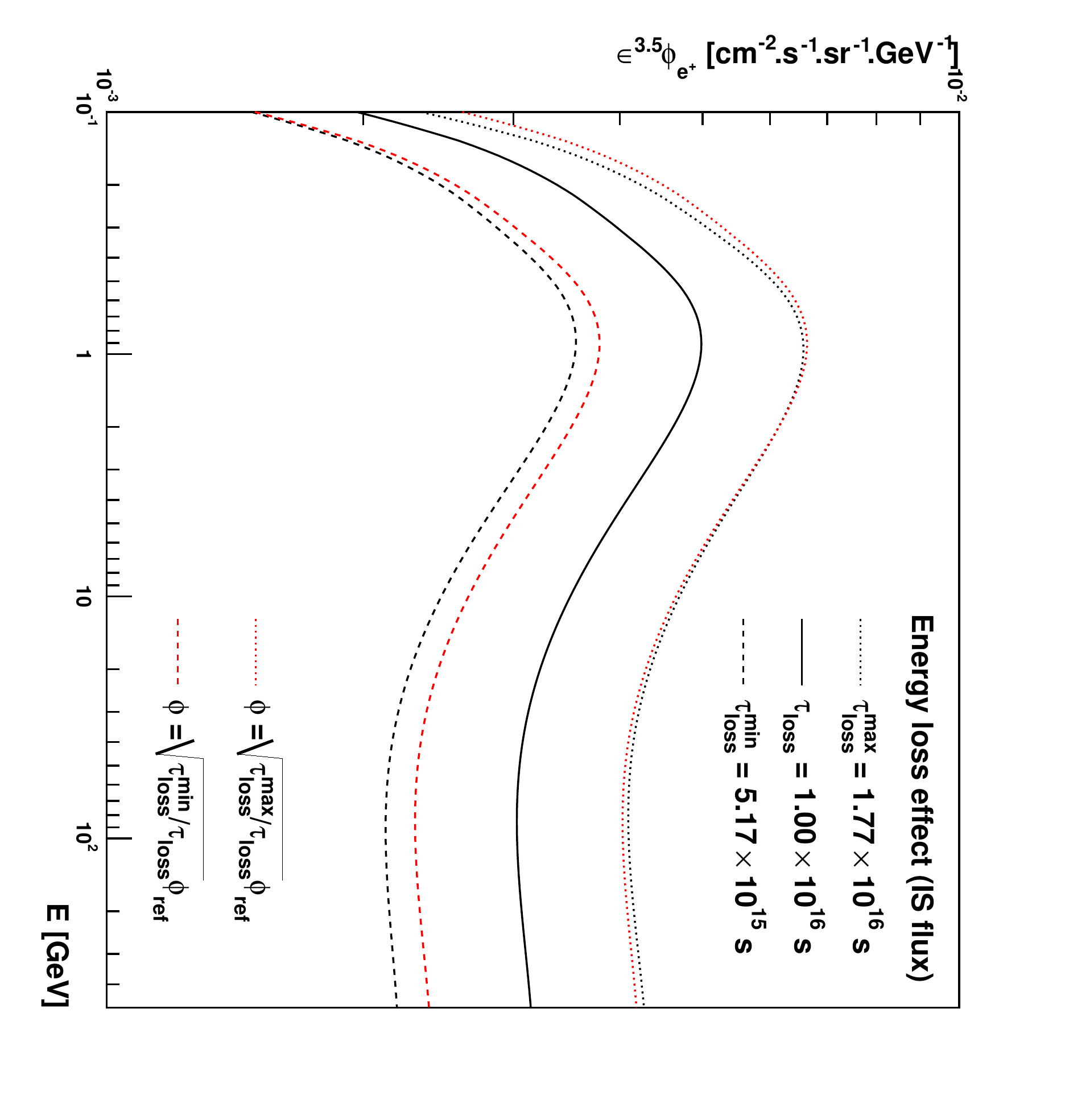}}
\caption{\label{fig:fig3-taue} Interstellar secondary positron flux 
  $E^{3.5} \Phi_{e^{+}}$ as a function of the energy at Earth, for different 
  values of the energy loss timescale $\tau_E$. The longer the timescale, 
  the larger the flux. The scaling relation $\phi\propto \sqrt{\tau_E}$ is 
  also reported.}
\end{center}
\end{figure}

The leading term caused by IC scattering by the ISRF (star and dust light) is is clearly affect by the most significant uncertainty. 
By combining all the contributions, we derive an uncertainty range for the energy--loss timescale of $\tau_E \in [5.17 \times 10^{15} ; 1.77 \times 10^{16} ]$~s.
As shown in Fig. \ref{fig:fig3-taue}, the uncertainty caused by $\tau_E$ does not exhibit a strong energy dependence. This can be understood from the expression of the positron Green function given in Eqs. (\ref{eq:pseudo-propag}) and (\ref{eq:propa}), and from the fact that the source term of secondaries is close to be homogeneously distributed in the thin Galactic disk. 
From this, the secondary positron flux roughly scales as $\sqrt{\tau_E}$, as also illustrated in Fig. \ref{fig:fig3-taue}. 
Since $\tau_E$ determines the positron propagation scale, the related uncertainties should 
be far stronger effeact on the primary contributions with a far greater spatial (or time) dependency, such as for instance nearby astrophysical or exotic point sources.
In the following, we have adopted as a standard the value of $\tau_E = 1 \times 10^{16}$~s.

We note that $\tau_{\textnormal{sync}}$ and $\tau_\star$ are not expected to remain constant throughout the entire diffusive halo (\cite{2000ApJ...537..763S}).
However, as we discuss extensively in Sect. \ref{sect:spatial}, most of the positrons detected at Earth with energies greater than a few GeV, are typically of local origin.
This is illustrated in Fig. \ref{fig:fig7}, where $\tau_E = 1\times 10^{16}$~s, for detected positrons of energy higher than 1~GeV: more than 75\% of the signal has originated within a distance of 2~kpc; the higher the detected energy, the higher this percentage.
Since we are not interested in very low energy positrons, we can safely neglect spatial 
variations in the magnetic field $B$ and the ISRF energy density $U_{\textnormal{rad}}$ discussed above. 

%--------------------------------------------------------------
\section{The positron flux and its uncertainties}
\label{sec:positron_flux_uncertainties}

\begin{figure}[t]
\begin{center}
\resizebox{\hsize}{!}{\includegraphics[width=\columnwidth]{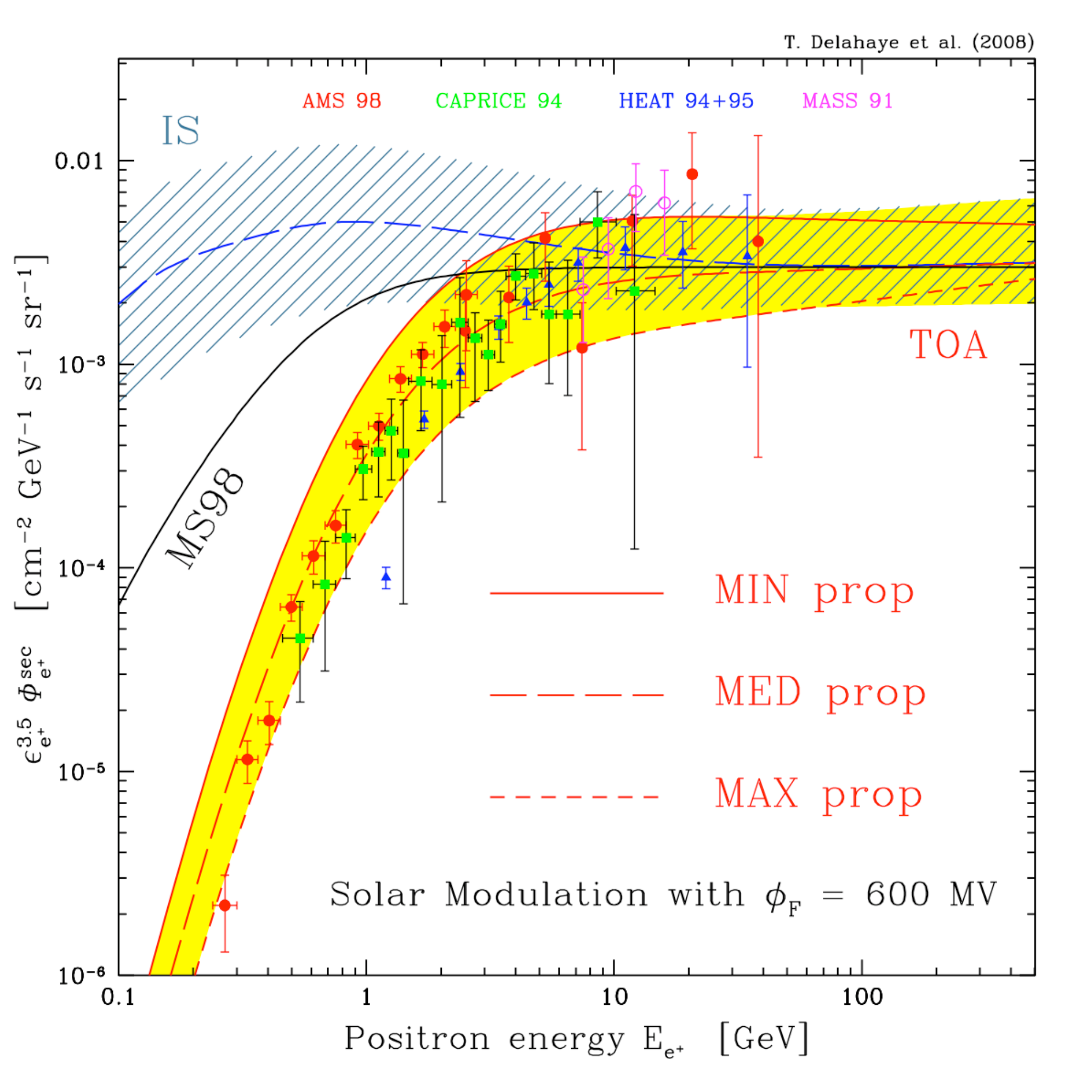}}
\caption{
Secondary positron flux as a function of the positron energy. The blue hatched 
band corresponds to the CR propagation uncertainty in the IS prediction, 
whereas the yellow strip refers to TOA fluxes.
The long--dashed curves feature our reference model with the \cite{Kamae2006}
parameterization of nuclear cross--sections, the \cite{bess_shikaze_etal_07} injection
proton and helium spectra and the MED set of propagation parameters.
The MIN, MED and MAX propagation parameters are displayed in 
Tab.~\ref{table:prop}.
%
%Data are taken from the Caprice, HEAT, AMS and MASS experiments.}
Data are taken from CAPRICE \citep{2000ApJ...532..653B}, HEAT \citep{1997ApJ...482L.191B}, 
AMS \citep{Aguilar:2007yf,2000PhLB..484...10A} and MASS \citep{2002A&A...392..287G}.}
\label{fig:fig3}
\end{center}
\end{figure}

Figure~\ref{fig:fig3} displays the calculated secondary positron flux modulated 
at solar minimum along with the most recent experimental data. We used a 
Fisk potential $\phi=600$ MV as applied in \cite{perko}. The MIN, MED and MAX cases 
are illustrated by the red solid, long--dashed and short--dashed 
lines, respectively, while the yellow area denotes the uncertainty in the propagated 
flux caused by the uncertainty in the astrophysical parameters. The nuclear 
cross--section from \cite{Kamae2006} and the \cite{bess_shikaze_etal_07} proton and helium fluxes were used.
We adopted the MED prediction as our reference model.
In the same figure, we also plot the interstellar flux. 
The upper long--dashed curve corresponds to the MED case whereas the 
slanted band indicates the uncertainty in the Galactic propagation parameters. The
solid line shows the IS flux from~\cite{1998ApJ...493..694M}.
Below $\sim$ 100 GeV, the yellow uncertainty band is delineated by the MIN 
and MAX models. As discussed at the end of the previous section, the MIN (MAX) 
set of parameters yields the highest (lowest respectively)
values for the secondary positron flux.
Since we considered more than about 1,600 different configurations
  compatible with the B/C ratio \citep{2001ApJ...555..585M}, other 
propagation models become important in determining the 
extremes of the uncertainty band at higher energies. 
The maximal flux at energies above 100 GeV
does not correspond to any specific set of propagation parameters over the 
whole range of energies, as already noted in \cite{2008PhRvD..77f3527D},where the 
case of positrons produced by dark matter annihilation was studied.

From Fig.~\ref{fig:fig3}, we see that the variation in the propagation parameters
induces an uncertainty in the positron flux, which reaches about one order of 
magnitude over the entire energy range considered here. It is a factor of 6 at 
1 GeV, and smoothly decreases down to a factor of 4 or less for energies 
higher than 100 GeV.
The agreement with experimental data is quite good at all energies within the
uncertainty band. The \cite{1998ApJ...493..694M} prediction of the IS secondary
positron flux as parameterized by \cite{1999PhRvD..59b3511B} is indicated by the
black solid curve, and hardly differs from our reference model (long--dashed
curve and MED propagation) above a few GeV.
The HEAT data points are in good statistical agreement with this MED model.

\begin{figure*}[t]
\begin{center}
\resizebox{\hsize}{!}{\includegraphics[width=\columnwidth]{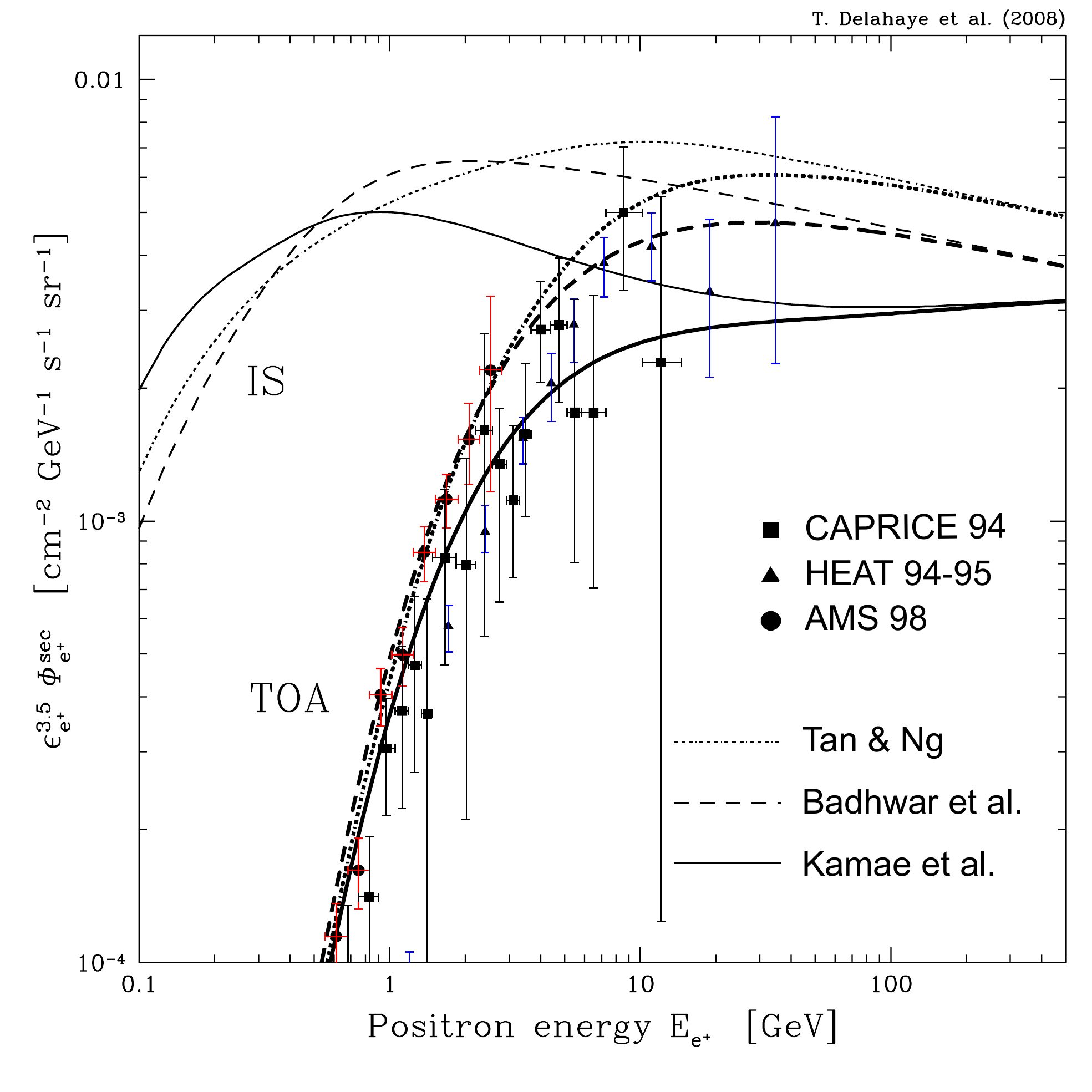}\includegraphics[width=\columnwidth]{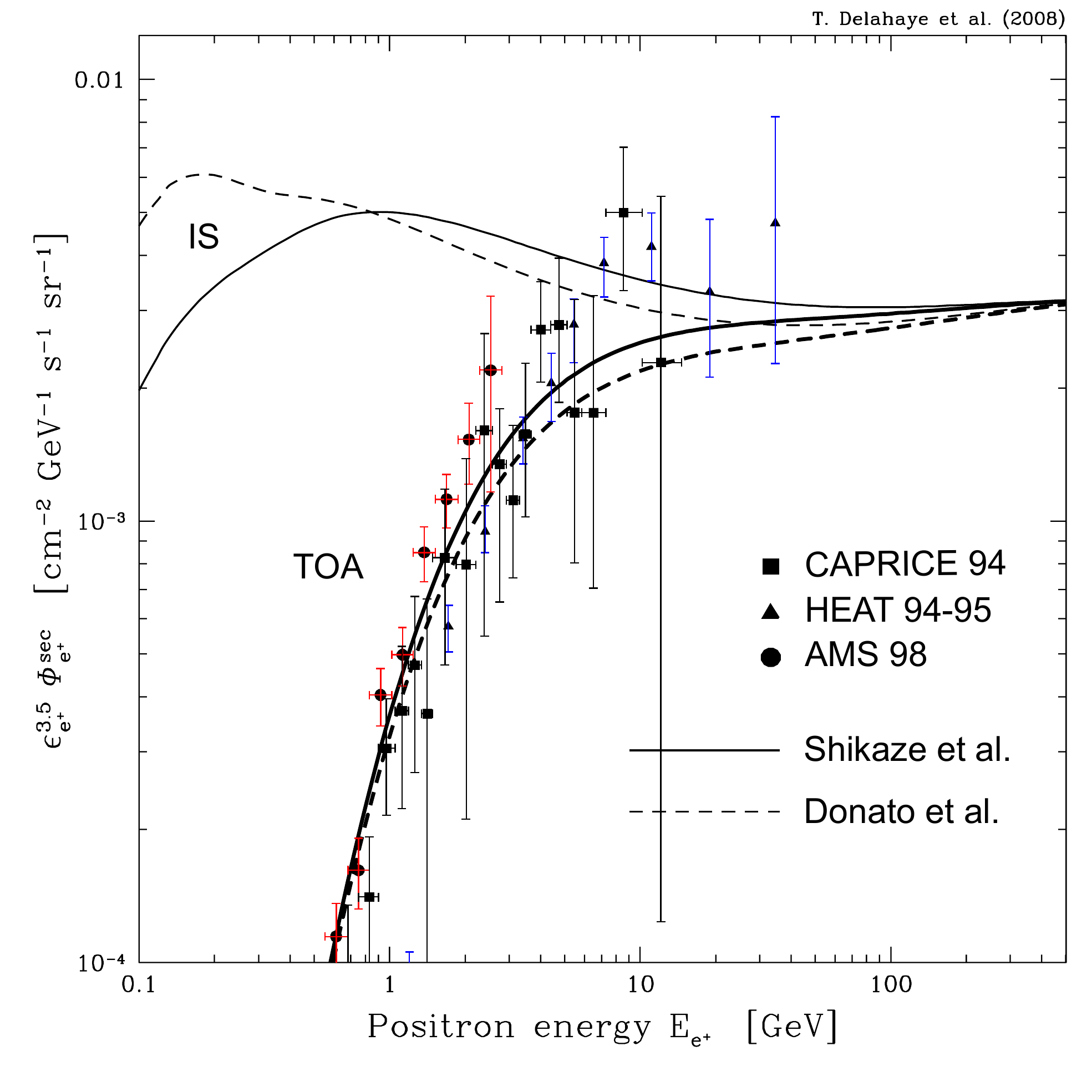}}
\caption{
{\sl Left:} TOA and IS positron spectra for three different nuclear cross--section parameterizations: Kamae (solid), Badhwar (dashed) and Tan \& Ng 
(dotted).
{\sl Right:} TOA and IS positron spectra for two different proton fluxes:
Shikaze (solid) and Donato (dashed).
In all cases, diffusion parameters are set to the MED case of
Tab.~\ref{table:prop}.
}  
\label{fig:fig4a}
\end{center}
\end{figure*}

The effects induced by different parameterizations of the nuclear production 
cross--sections and by the variation in the proton injection spectrum are shown in 
Fig.~\ref{fig:fig4a}. In the left panel, we present the TOA positron fluxes 
calculated from the \cite{tan_ng_1983_b} (dotted), \cite{badhwar_1977} 
(dashed), and \cite{Kamae2006} (solid) cross--section models for the MED 
propagation scheme and the \cite{bess_shikaze_etal_07} proton and helium injection 
spectra. The \cite{Kamae2006} model leads systematically to the lowest flux. 
For positron energies $\lsim$ 1 GeV, the three cross--section parameterizations 
differ by just a few percent, while the differences are significantly larger at 
higher energies. Figure~\ref{fig:fig4a} translates the 
uncertainties in the source term $q_{e^{+}}$ featured in 
Fig.~\ref{fig:p_and_alpha_spectra}. Consequently, the flux obtained at 10 GeV
with the \cite{tan_ng_1983_b} or \cite{badhwar_1977} parameterization
is a factor of 2 or 1.6, respectively, 
higher than the reference case \citep{Kamae2006}. This trend is 
confirmed at higher energies, although the differences between the various 
models are smaller above 200~GeV.

The uncertainties caused by the proton and helium spectrum parameterizations 
are the least relevant to this analysis. This is demonstrated in the right panel of
Fig.~\ref{fig:fig4a}, where we compare the positron flux of the reference model
(solid lines) with the flux obtained when the \cite{bess_shikaze_etal_07}
parameterization of the incident spectra was replaced by that of
\cite{2001ApJ...563..172D}. The differences are at most 10--15\% around 10 GeV,
and are negligible in the lower and higher energy tails.

\subsection{Spatial origin of the positrons}
\label{sect:spatial}
\begin{figure}[t]
\begin{center}
\resizebox{\hsize}{!}{\includegraphics[width=\columnwidth]{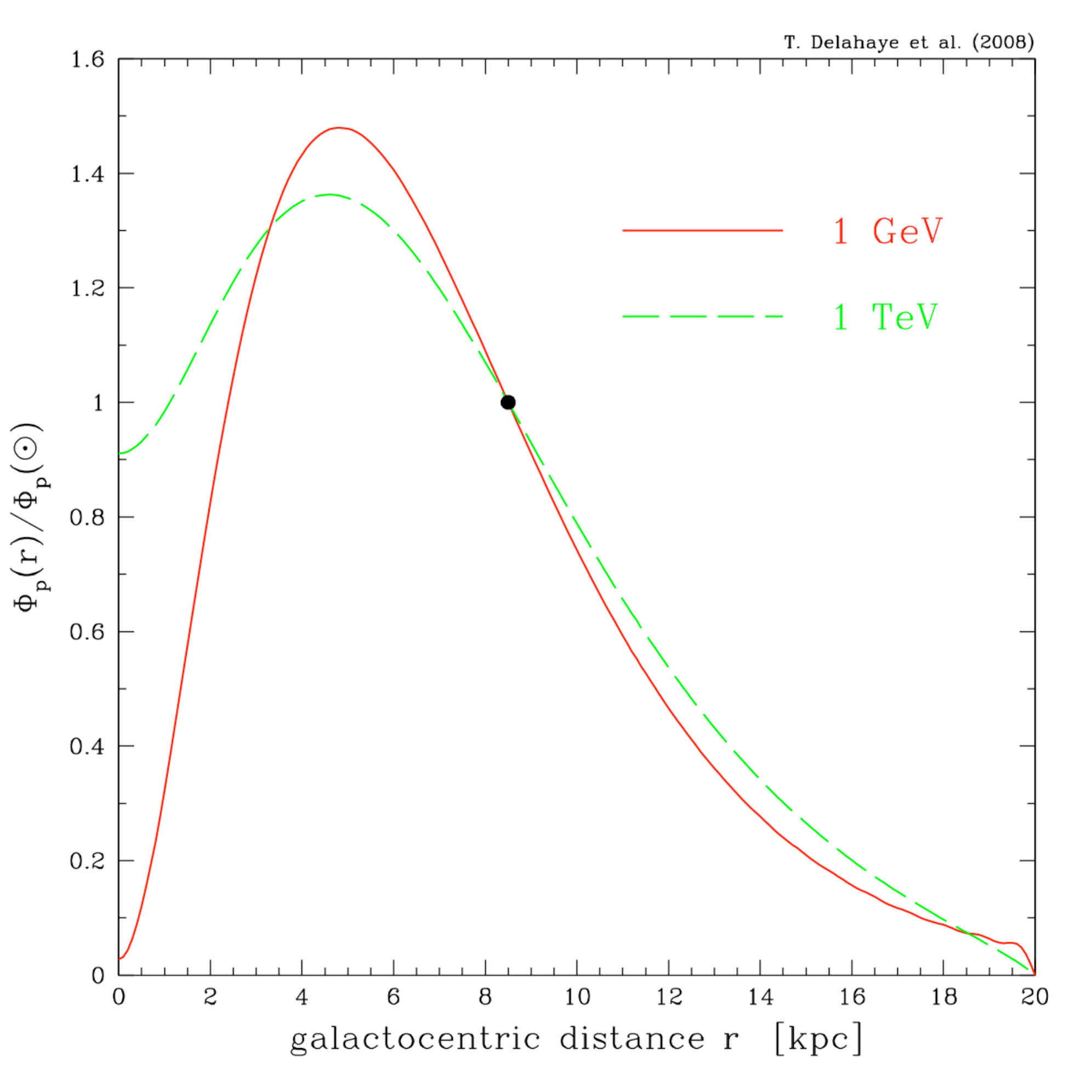}}
\caption{
Ratio of the proton flux at radius $r$ to the solar value from 
\cite{bess_shikaze_etal_07}.
In this plot, retro--propagation has been taken into account, and all 
propagation effects of the MED configuration (\ie, convective wind, spallation, 
and diffusion) have been included.
The dot refers to the Solar System position in the Galaxy.} 
\label{fig:fig6a}
\end{center}
\end{figure}

At every location in the Galaxy, the positron production by spallation
is determined by the local flux of cosmic ray proton and helium projectiles.
Their spatial distribution $\Phi(\mathbf{x},E)$ was assumed to be 
constant and set equal to
the value $\Phi_{\odot}(E)$ measured at the Solar System location.
However, we note that these CR primaries also diffuse in the Milky Way, so that
their flux should exhibit a spatial dependence. The positron source term 
$q_{e^{+}}$ should vary accordingly inside the Galactic disk. The behavior of 
the proton and helium fluxes with radius $r$ can be inferred readily from 
their measured values $\Phi_{\odot}(E)$ once the propagation parameters are 
selected. This so--called retro--propagation was implemented in 
the original B/C analysis by \cite{2001ApJ...555..585M}.
The radial variation in the proton flux is presented in Fig.~\ref{fig:fig6a}
for two quite different proton energies, and is found to be significant. This is
why we questioned the hypothesis of a homogeneous positron production
throughout the disk and found nevertheless that it remains viable in spite
of the strong radial dependence of $\Phi(r,E)$.

\begin{figure}[t]
\begin{center}
\resizebox{\hsize}{!}{\includegraphics[width=\columnwidth]{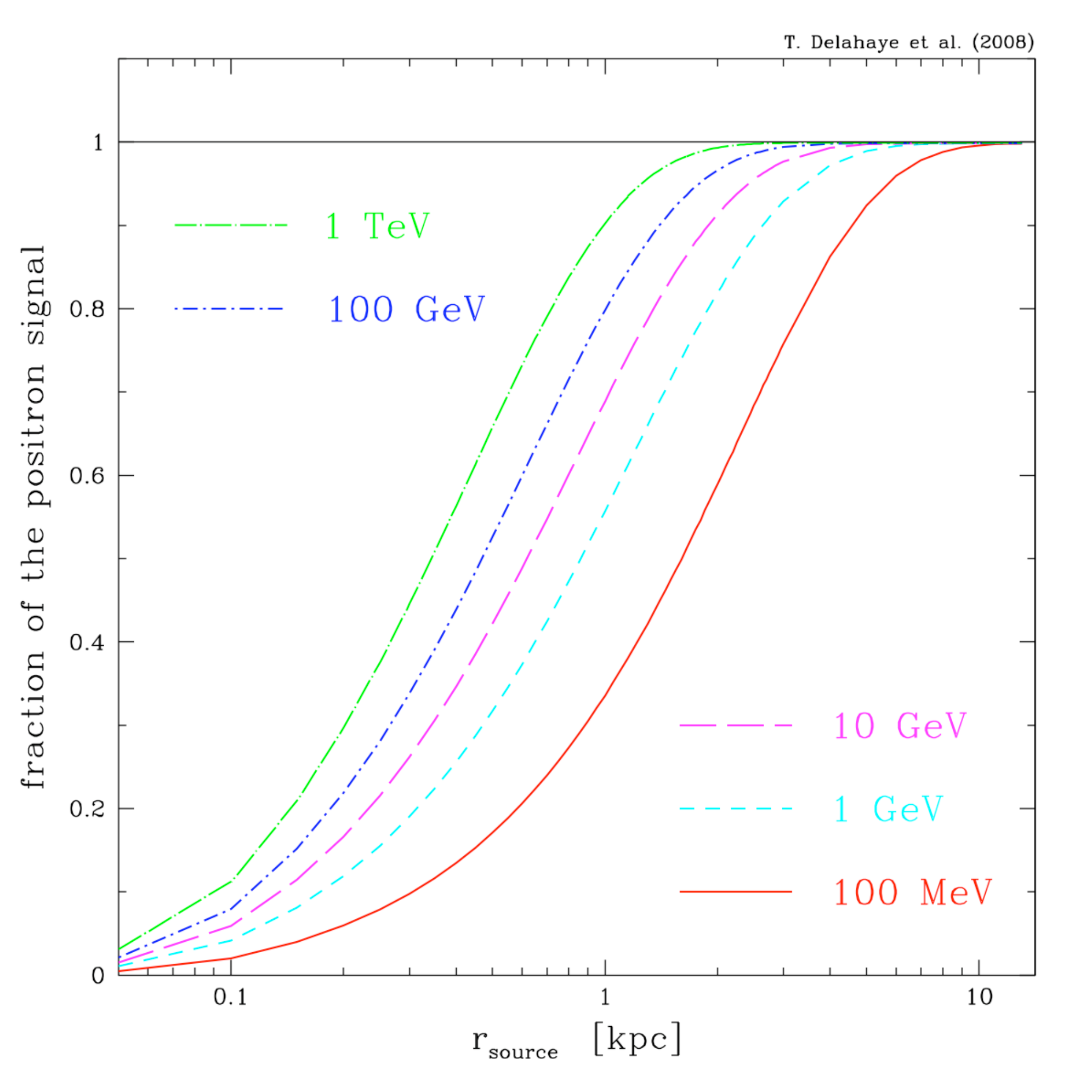}}
\caption{
Fraction of the positrons detected at the Earth which are produced
within a disk of radius $r_{\rm source}$. The larger the energy, the
closer the source.} 
\label{fig:fig5}
\end{center}
\end{figure}

\begin{figure}[t]
\begin{center}
\resizebox{\hsize}{!}{\includegraphics[width=\columnwidth]{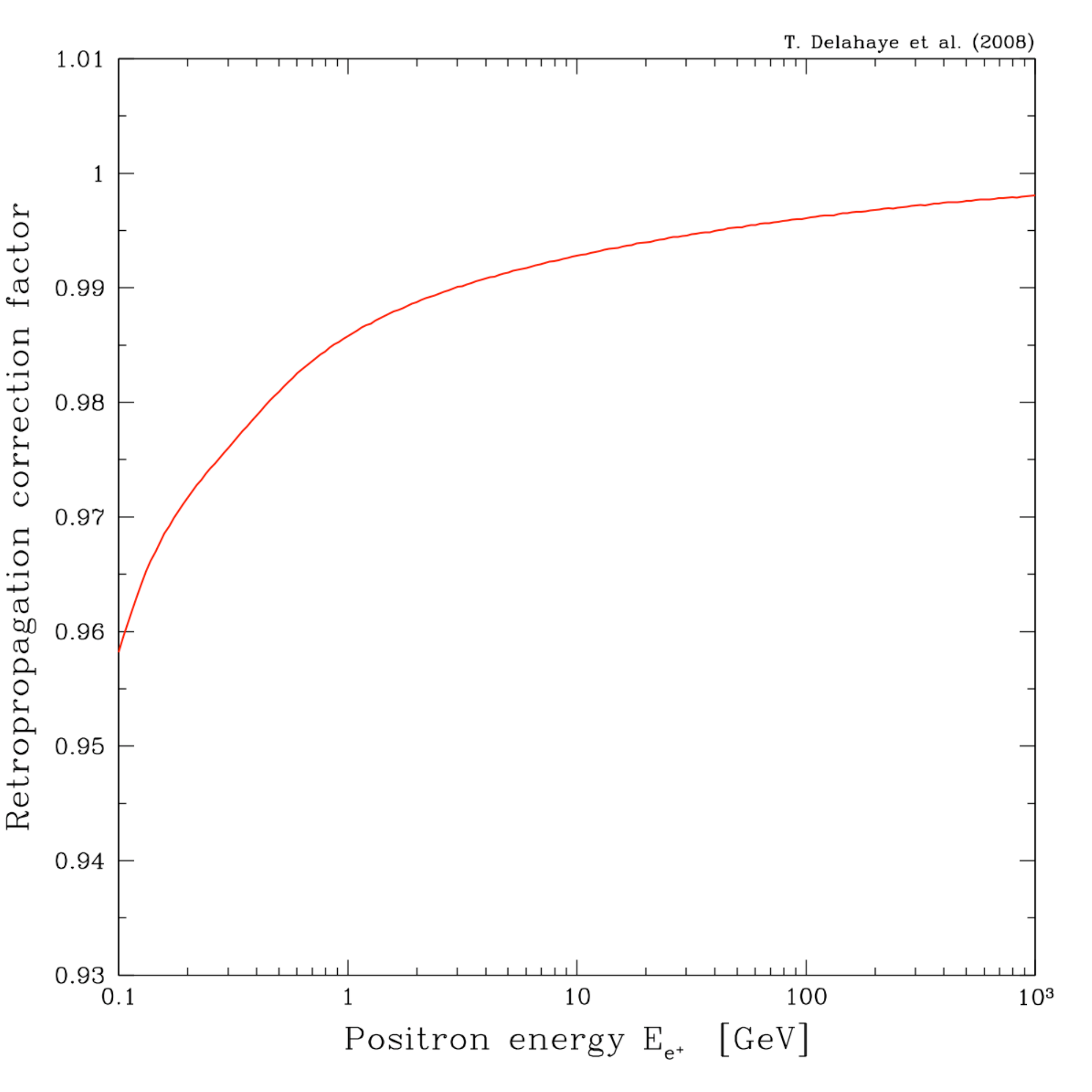}}
\caption{
Ratio of the positron flux computed with and without 
retro--propagation, as a function of positron energy.} 
\label{fig:fig6b}
\end{center}
\end{figure}

This is because positrons reaching the Earth were mostly 
created locally, \ie, in a region in which the proton flux does not
differ significantly from the local value $\Phi_{\odot}(E)$.
We evaluated the contribution to the total
signal from a disk of radius $r_{\rm source}$ surrounding the Earth,
which was modeled with the source term
\begin{equation}
q_{\rm \, source}(r,E) = q_{e^{+}}(r,E) \times \Theta(r_{\rm source} - r),
\end{equation}
where $\Theta(x)$ is the Heaviside function and $r$ measures the radial
distance from the Solar System.
The positron flux yielded by $q_{\rm \, source}$ is
$\phi_{e^{+}}^{\odot}(r_{\rm source},E)$, whose contribution to the total signal
$\phi_{e^{+}}^{\odot}(E)$ is plotted in Fig.~\ref{fig:fig5} as a function of
$r_{\rm source}$, for several values of the positron energy $E$.
Most of the positron signal originates at short distances, especially at high
energy. At 1~TeV, more than 80\% of the positrons are created
within 1~kpc while more than half of the 100 MeV positrons come from less
than 2~kpc. Energy losses are indeed quite efficient. They reduce
the positron horizon strongly as the energy increases.
This is why the CR proton and helium fluxes are close to their solar
values when averaged over the positron horizon scale.
Taking the retro--propagation of projectile spectra into account
has therefore little effect on the positron flux, as is clear in
Fig.~\ref{fig:fig6b}.

%\subsection{Diffusive reacceleration}
\subsection{Diffusive reacceleration and full energy losses}
\label{sec:diffusive_reacceleration}

Space diffusion and energy losses through inverse Compton scattering
and synchrotron emission were the only processes that we had considered.
We had neglected many other mechanisms that may also affect
positron propagation. Galactic convection can sweep cosmic rays out of
the diffusive halo and is associated with adiabatic energy losses. The
drift of the magnetic turbulent field with respect to the Galactic
frame with velocity $V_{a}$ induces both a diffusion in energy space
and a reacceleration of particles. This so--called diffusive
reacceleration was discussed in the original analysis
of \cite{1998ApJ...493..694M}, but it does not appear in the fitting
formula proposed by \cite{1999PhRvD..59b3511B}. Finally, bremsstrahlung
and ionization should come into play at low energies, below $\sim$ 1~GeV.

\begin{figure*}[t]
\begin{center}
\resizebox{\hsize}{!}{\includegraphics[width=\columnwidth]{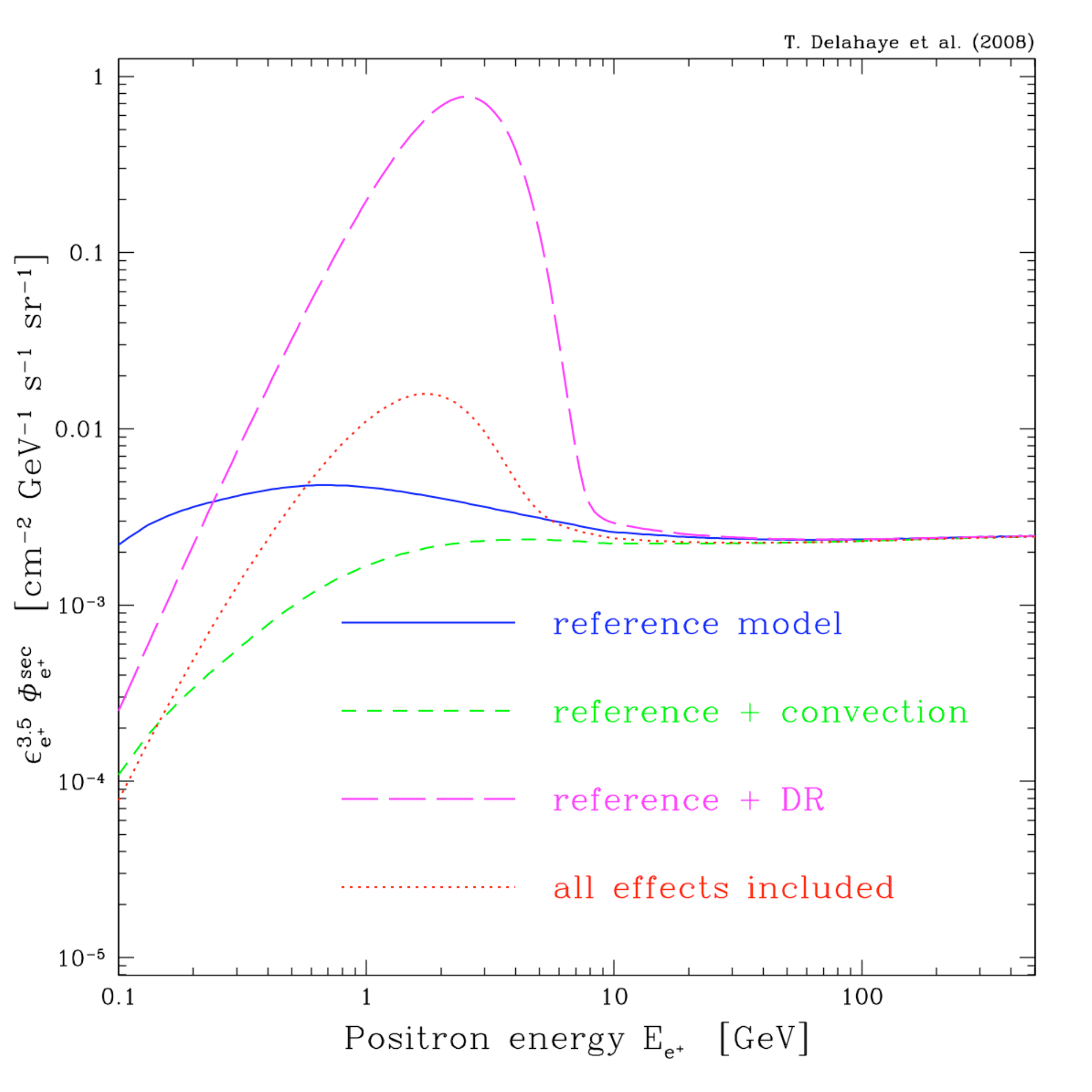}\includegraphics[width=\columnwidth]{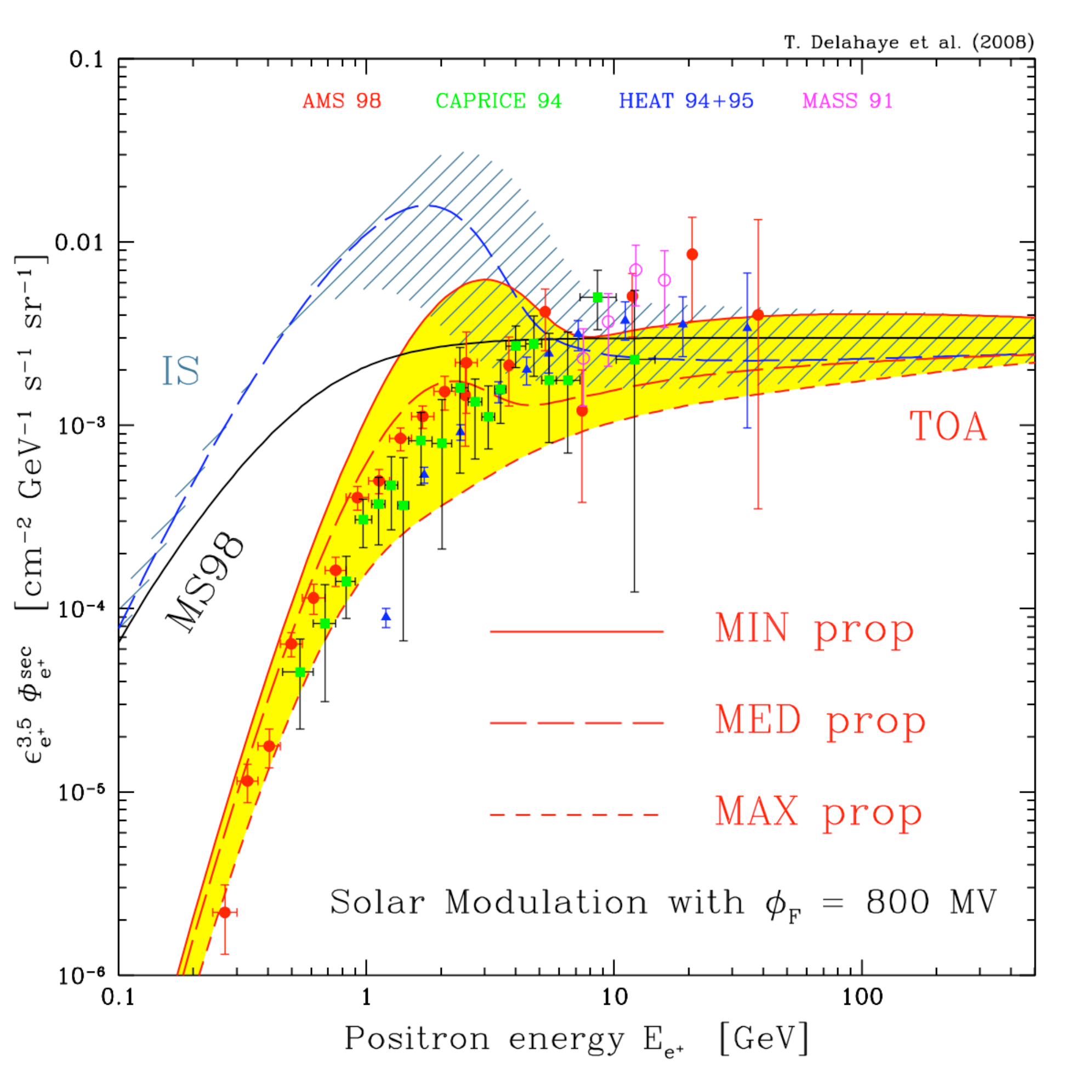}}
\caption{
{\sl Left panel:} The reference model of 
Sect.~\ref{sec:positron_flux_uncertainties} is featured here with various 
effects turned on or off. Space diffusion and energy losses from IC and 
synchrotron emission lead to the solid curve.
When diffusive reacceleration is added, we get the long--dashed line and its
spectacular bump around 3~GeV.
The short--dashed curve is obtained by replacing diffusive reacceleration by
galactic convection. The spectrum becomes depleted at low energies.
Including all the processes yield the dotted line. Diffusive reacceleration and
convection are both relevant below a few GeV and induce opposite effects.
{\sl Right panel:}
The hatched blue (IS) and yellow (TOA) regions of Fig.~\ref{fig:fig3} delineated
by the MAX and MIN curves are featured here with all the effects included.
Above a few GeV, we get the same results as before.
Data are taken from CAPRICE \citep{2000ApJ...532..653B}, HEAT \citep{1997ApJ...482L.191B}, 
AMS \citep{Aguilar:2007yf,2000PhLB..484...10A} and MASS \citep{2002A&A...392..287G}.
} 
\label{fig:fig7}
\end{center}
\end{figure*}

We note that once all these effects are considered, there is no longer an
analytical solution to the diffusion equation, at least none that we know.
Synchrotron and IC energy losses occur all over the diffusive halo,
whereas the other energy-loss mechanisms as well as diffusive reacceleration
are localized within the Galactic plane. A completely numerical approach
is always possible of course -- dealing for instance with a realistic distribution of gas, magnetic fields, and 
interstellar radiation fields -- but this has never been our philosophy so
far. Keeping calculations as analytical as possible has always been
our guiding principle.
We have therefore derived an approximate solution to the complete
diffusion equation where the IC and synchrotron losses have been suitably
renormalized and assumed to take place only in the disk. Obviously this is
not fully correct, but a close inspection of the left panel of
Fig.~\ref{fig:fig7} and its solid curve indicates these values
agree closely with
the reference model featured in Fig.~\ref{fig:fig3} by the long--dashed
blue line. The diffusion equation may now be expressed as:
\begin{eqnarray}
\mathbf{\nabla} \! \cdot
\left\{
- \, K_0 \, \epsilon^{\delta} \, \mathbf{\nabla} N \, + \,
\mathbf{V}_{C}(z) N
\right\} \; & + & \\
+ \, 2 \, h \, \delta(z) {\displaystyle \frac{\partial}{\partial \epsilon}}
\left\{
b^{\rm loss}(\epsilon) \, N \, - \,
K_{\epsilon \epsilon} \, \frac{\partial N}{\partial \epsilon}
\right\} & = & q_{e^{+}}(\mathbf{x},\epsilon) \;\; , \nonumber
\label{eq:master_1}
\end{eqnarray}
where $\epsilon = E/E_{0}$ and $E_{0} = 1$ GeV as defined in
Sect.~\ref{subsec:green_function}. The convective wind varies with
the vertical coordinate as:
\begin{equation}
\mathbf{V}_{C}(z) =
\begin{cases}
\;\; V_C \, \mathbf{u}_z & \text{if } z>0 \\
-    V_C \, \mathbf{u}_z & \text{if } z<0.\\
\end{cases}
\label{eq:convect}
\end{equation}
The energy-loss term is a combination of several contributions:
\begin{equation}
- b^{\rm loss}(\epsilon) =
\begin{cases}
\;\;\displaystyle \frac{\epsilon^2}{\tau_E} \hspace{0.8cm} \text{inverse 
Compton and synchrotron} \\
\displaystyle + {\mathbf{\nabla} \cdot \mathbf{V}_{C}} \, {\frac{p^{2}}{6 h \epsilon}}
\hspace{2.2cm} \text{adiabatic losses} \\
\displaystyle + K_{b} \, n_H \, \epsilon \hspace{2.8cm} \text{bremsstrahlung}\\
\displaystyle + K_{i} \, n_H \,
\left\{ 3 \ln \left( \frac{E}{m_e} \right) + 19.8 \right\}
\hspace{0.4cm} \text{ionisation.}
\end{cases}
\label{eq:losses}
\end{equation}
The values of the constants $K_{b}$ and $K_{i}$ can be found in 
\cite{1994hea..book.....L}.
Diffusive reacceleration is taken care of by using a coefficient
\begin{equation}
K_{\epsilon \epsilon} = \frac{2}{9} \,
{\displaystyle \frac{V_{a}^{2}}{K_{0}}} \, \epsilon^{2 - \delta}\;\; .
\end{equation}
The positron density $N(r,z,\epsilon)$ is Bessel expanded. The coefficients
$N_{i}(z,\epsilon)$ are determined all over the diffusive halo
except for the
normalizations $N_{i}(0,\epsilon)$. Each of these normalizations satisfies a diffusion
equation in energy space, which we solve using a Crank--Nicholson semi--implicit
method.

%Discussion of the left panel of Fig.~\ref{fig:fig7}
The solid line of Fig.~\ref{fig:fig7} considers only space diffusion and energy
losses by IC scattering and synchrotron emission.
When these processes are supplemented by diffusive reacceleration,
we derive the long--dashed curve with a noticeable bump at $\sim$ 3~GeV.
Below that value, positrons are reaccelerated and their energy
spectrum is shifted to higher energies. Above a few GeV, IC scattering
and synchrotron emission dominate over diffusive reacceleration,
inducing a shift in the spectrum towards lower energies. Positrons accumulate
in an energy region where energy losses and diffusive reacceleration
compensate each other, hence a visible bump which is already present in
the analysis by \cite{1998ApJ...493..694M}.
The short--dashed line is obtained by replacing diffusive reacceleration by
Galactic convection. The wind is active at low energies, where space
diffusion is slow. Positrons are drifted away from the Galaxy and their
flux at the Earth is depleted.
We note that diffusive reacceleration and Galactic convection were
included separately by \cite{2005JCAP...09..010L} in their prediction of the
positron spectrum, with the net result of either overshooting (diffusive
reacceleration) or undershooting (galactic convection) the data.
If we now incorporate both processes and add the various
energy-loss mechanisms, we derive the dotted curve, which also
contains a bump, although of far smaller amplitude. The bump cannot be distinguished
from the solid line for energies above a few GeV.
This is the energy region where dark matter species are expected to distort the 
secondary spectrum and a calculation based solely on space diffusion and 
energy losses from IC scattering and synchrotron emission is perfectly safe.

Below a few GeV, the situation becomes more complicated, several effects
at stake modifying the blue hatched IS and yellow TOA uncertainty
intervals in
Fig.~\ref{fig:fig3} as displayed in the right panel of Fig.~\ref{fig:fig7}.
Reproducing the GeV bump of the IS flux with the data now requires a higher Fisk
potential of 850 MV. The agreement seems reasonable below a few GeV, although
a more detailed investigation would require a refined solar modulation model.

%--------------------------------------------------------------
\section{The positron fraction}
\label{sec:positron_fraction}

\begin{figure}[t]
\begin{center}
\resizebox{\hsize}{!}{\includegraphics[angle=90,width=\columnwidth]{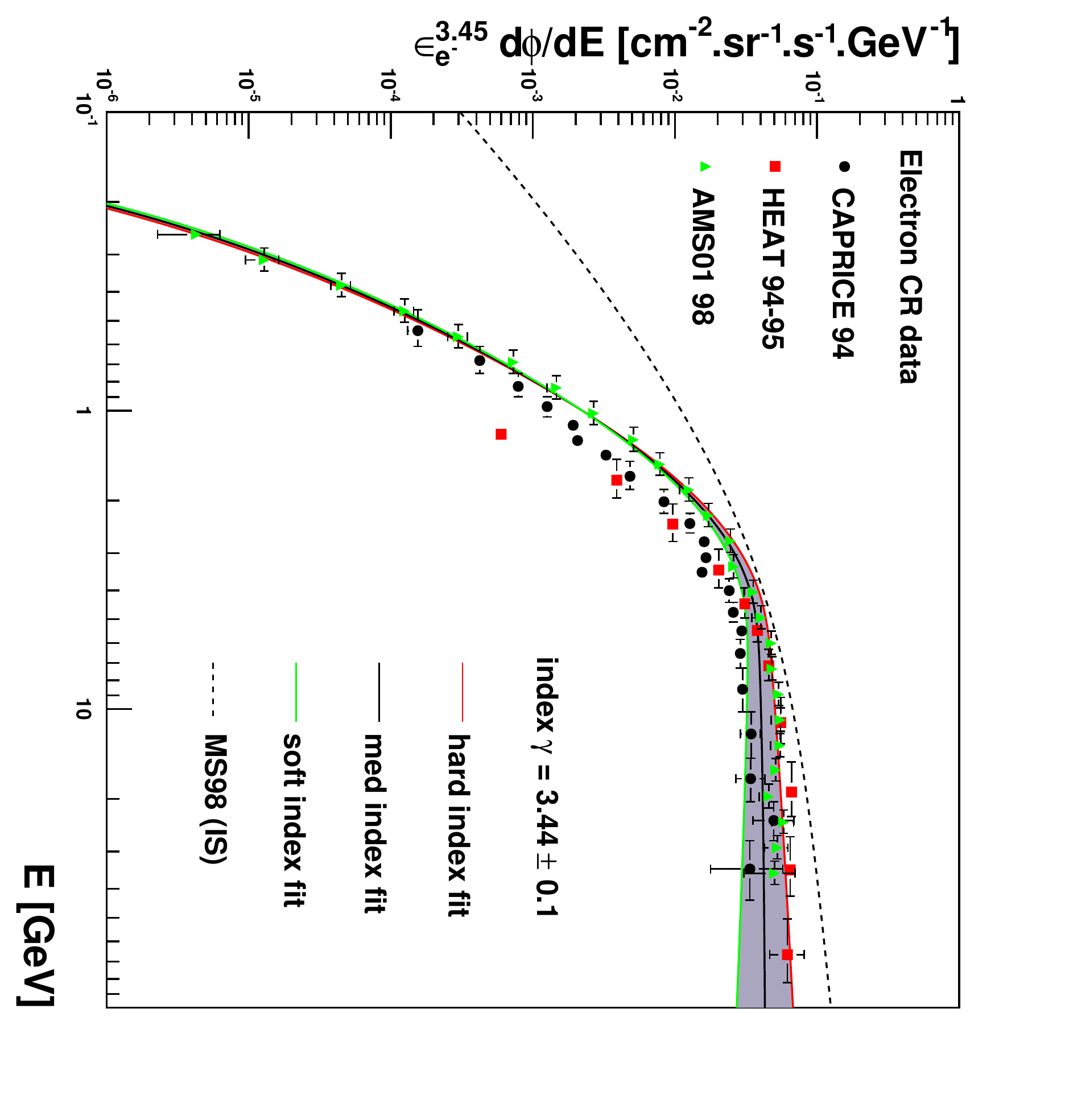}}
\caption{Electron flux parameterization.} 
\label{fig:fig_electron_fit}
\end{center}
\end{figure}

\begin{figure*}[t]
\begin{center}
\resizebox{\hsize}{!}{\includegraphics[width=\columnwidth]{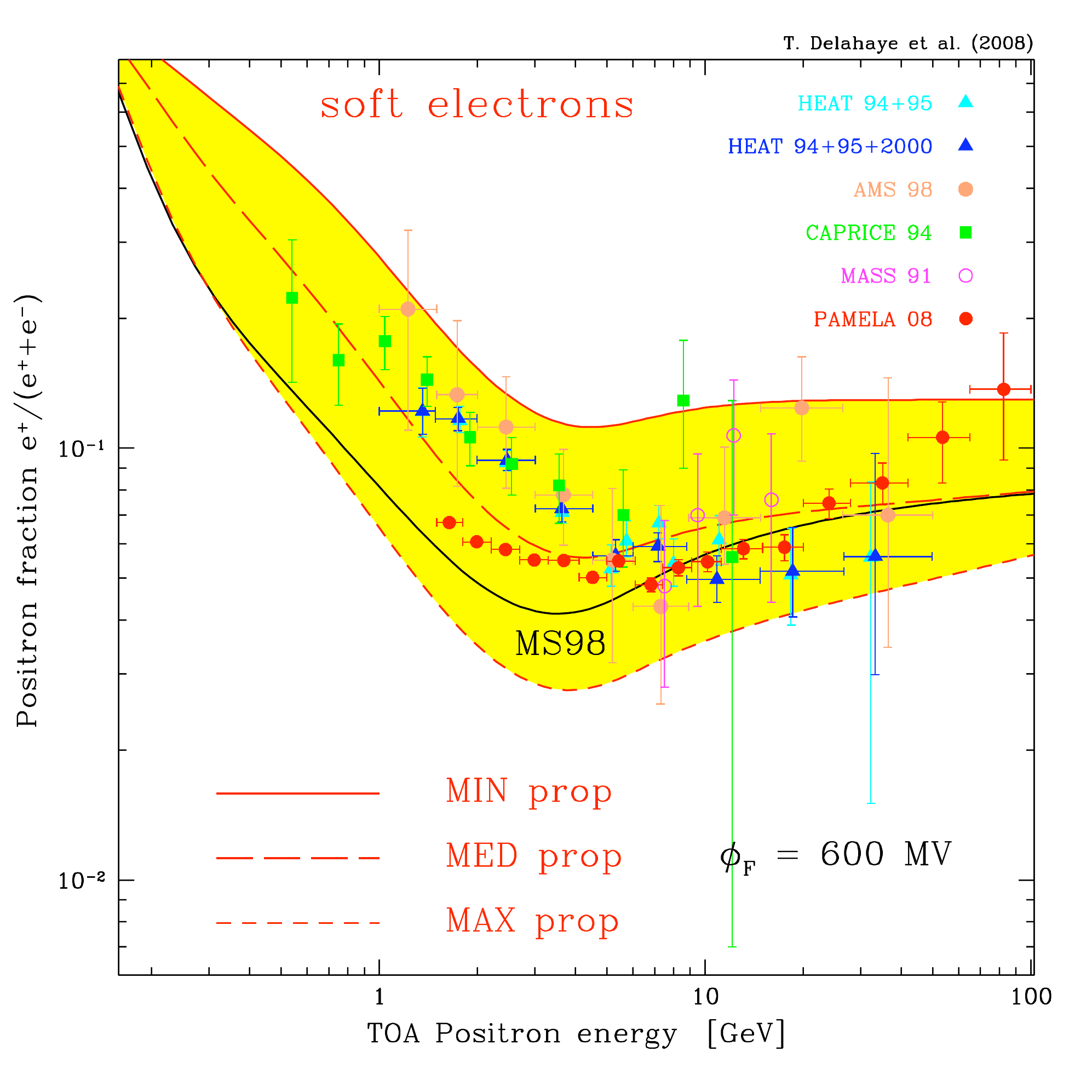}\includegraphics[width=\columnwidth]{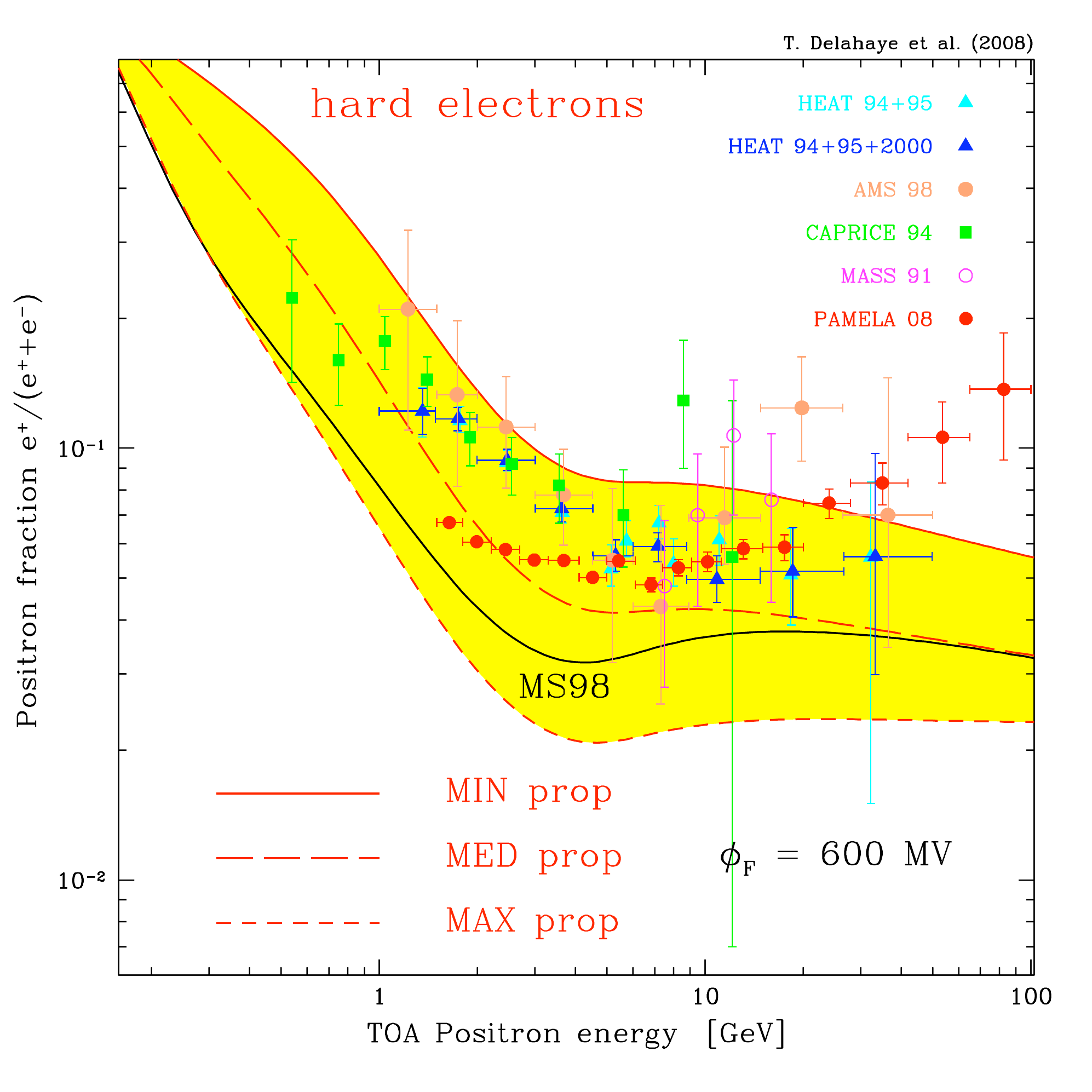}}
\caption{Positron fraction as a function of the positron energy, for
a soft (left panel) and hard (right panel) electron spectrum. 
Data are taken from CAPRICE \citep{2000ApJ...532..653B}, HEAT \citep{1997ApJ...482L.191B}, 
AMS \citep{Aguilar:2007yf,2000PhLB..484...10A}, MASS \citep{2002A&A...392..287G} and PAMELA \citep{PAMELA08}.} 
\label{fig:pf-1}
\end{center}
\end{figure*}

The question of whether a positron excess is being observed in cosmic ray 
measurements remains after many years, and is usually addressed in 
terms of the the so--called \emph{positron fraction}, \ie, the quantity 
$\phi_{e^+}/(\phi_{e^+} + \phi_{e^-})$. This excess was pointed out 
by~\cite{1998ApJ...493..694M} when they derived their predictions of both the 
secondary positron flux and the primary plus secondary electron fluxes. This 
led to many possible interpretations, such e.g., a potential positron 
injection from dark matter annihilation in the 
Galaxy~\citep[e.g.,][]{1999PhRvD..59b3511B} or from 
pulsars~\citep[e.g.,][]{1995A&A...294L..41A,1989ApJ...342..807B,2007A&A...474..339G}.

The PAMELA collaboration reported the positron fraction from 1.5 to 
100 GeV with unprecedented statistics quality \citep{PAMELA08}. The PAMELA data 
are reported in Fig. \ref{fig:pf-1}, which we further explain in this 
section. Compared to the typical reference prediction 
by~\cite{1998ApJ...493..694M}, a large excess appears. Our new 
predictions of the secondary positron flux and its theoretical uncertainties 
allow us to discuss in greater depth the interpretation of the excess positron 
fraction. We note first that we have shown in the previous sections that our 
predictions are consistent with the available positron measurements, 
excluding the PAMELA data since only their positron 
fraction data have so far become public. From these pre-PAMELA data, we therefore remark 
that an excess is hardly observed when considering the positron measurements only.

One of two crucial ingredients needed to derive the positron 
fraction is of course the electron flux. It was already noted 
by~\cite{1998ApJ...493..694M} that a change in the electron spectrum
may affect the existence of an excess in the positron 
fraction. These authors compared the positron fraction obtained 
when using their own prediction of the electron spectrum, with that obtained 
when using the prediction of~\cite{1982ApJ...254..391P} based on the leaky 
box propagation model, and hence illustrated the difference due to the 
electrons. Today, we can attempt to take advantage of the higher quality existing 
electron data, and complement them with our theoretical calculation of the 
positron flux and its uncertainties.

It is nevertheless difficult to constrain the electron flux using
the existing data, because they are available not only for limited energy 
ranges but also exhibit some differences in the absolute 
normalizations as well as the spectral shapes. Furthermore, the 
measurements available are mainly for energies below $\sim 50$ GeV. For instance, 
the AMS electron data points of \cite{Aguilar:2007yf} and 
\cite{2000PhLB..484...10A}, which are known to be among the most precise data 
sets to date, reach $\sim$ 30 GeV only. Other complications are 
expected from astrophysical modeling arguments. For instance, it is likely that 
different spectral contributions are important at different energies, such as 
secondary electrons at low energy, and primary electrons from local (distant) 
astrophysical cosmic ray sources at higher (intermediate, respectively) 
energies: this may imply that a blind fit to the data is not useful in the absence 
of any modeling insights. Although accounting for all these subtleties is 
beyond the scope of this paper, we can illustrate the importance of 
the electron spectrum by using a simple parameterization 
that is constrained reasonably up to $\sim$ 100 GeV, \ie, encompassing the PAMELA energy 
range, which fits the data at lower energies. 
We emphasize that 
predicting the electron flux is unnecessary to interpreting the positron 
contribution to the positron fraction as soon as electron data become
available. Using the electron data directly instead of a model is 
also likely to provide a far more accurate description of any excess.

Making use of the electron data, we have therefore modeled an electron 
spectrum by a power law $\propto E^{-\gamma}$ at energies above a few GeV and 
up to 100 GeV, and allowed a scatter in the spectral index to account for the 
dispersion in the different sets of data. Because the permitted 
range of spectral indices that provide good fits is significant, we have 
applied restrictions by taking the central value and variance found 
by~\cite{2004ApJ...612..262C}, \ie, $\gamma = 3.44\pm 0.03$. Thus, an 
index range defined by $\gamma\pm 3\sigma$ allows to encompass most of the 
available data below 100 GeV. The normalization of this electron spectrum as 
well as the low energy part have been adapted to the AMS data, 
which therefore includes the solar modulation effects. The choice of AMS is 
motivated by being likely to be the experimental setup least 
affected by systematic errors, but we emphasize that other setups 
are possible. Our 
parameterization of the electron flux is illustrated in 
Fig.~\ref{fig:fig_electron_fit}, where the data from HEAT, CAPRICE, and AMS are 
presented. The dispersion observed in the data above a few GeV and below 
$\sim$ 100 GeV, \ie, over the entire PAMELA energy range, is 
reproduced well by taking a spectral index of $\gamma = 3.44\pm 0.1$, as 
mentioned above. To interpret the positron 
fraction measurements, we therefore considered two cases for the electron 
flux, a \emph{soft} spectrum with index 3.53, and a \emph{hard} spectrum with 
index 3.35. Finally, we note that we have also presented in 
Fig.~\ref{fig:fig_electron_fit} the interstellar electron model of 
\cite{1998ApJ...493..694M}, which, while those authors have since considerably 
improved their model, is still widely used as a reference 
for predicting the positron fraction: we note that this model clearly 
overshoots the AMS data above a few GeV, and would therefore
underestimate the positron fraction, should the electron data of AMS 
be close to the true flux and unaffected by systematics or unknown transient effects.

In Fig.~\ref{fig:pf-1}, we show the positron fraction obtained for both the 
\emph{soft} (left panel) and \emph{hard} (right panel, respectively) electron 
spectra. For the positrons, we have used the Kamae nuclear cross--sections and 
the Shikaze proton and alpha injection spectra. The yellow band is bounded 
from below (above) by the MAX --- short dashed curve --- (MIN --- solid curve 
---, respectively) set of propagation parameters, while the central long-dashed 
curve represents the MED configuration. A solar modulation with $\phi=600$ MV 
has been applied to the positron flux, which corresponds to the level of solar 
activity during the data taking of AMS. In the same figure, we also report the 
positron fraction obtained with the positron flux of 
\cite{1998ApJ...493..694M}, but with our parameterization of the electron 
flux.

We see that, in the \emph{hard} index case, a sizeable excess is present in 
the high energy tail. The MED reference curve is marginally compatible with 
the HEAT and AMS data above 10--20 GeV, which instead lies closer to the upper 
border of our predictions, thus favoring the MIN model, which is
consistent with our predictions of the secondary positron flux 
(cf. Fig. \ref{fig:fig3}), should the positron 
flux be dominated by a single secondary contribution. Therefore, when the 
theoretical uncertainties are considered, a clear assessment of an excess is 
not statistically significant on the basis of the HEAT and AMS data alone, 
apart from the $2\sigma$ tension with the AMS data point at 12 GeV. 
Nevertheless, in the case of the PAMELA data, the MED reference flux is 
clearly incompatible with the experimental determinations for energies above 
10 GeV. Even when theoretical uncertainties in the positron flux are taken 
into account, an excess is probably present for a hard electron 
spectrum.

When using the \emph{soft} electron parameterization instead, we see that 
although an excess might still be apparent, its amplitude has strongly 
decreased, making it of least statistical relevance. The MED model
indeed reproduces all the data-sets well from a few GeV up to 40 GeV and a deviation is 
present for the last two bins of PAMELA, where the error bars are large due to 
reduced statistics.
The PAMELA data may therefore be indicative of
an excess also for a soft electron spectrum and energies 
above 50 GeV, but once the theoretical uncertainties on secondary positrons 
and statistical fluctuations in the data are taken into account, the amplitude 
of this excess is of least relevance. This implies that, for a soft electron 
spectrum, the secondary positron yield might still represent a very important 
contribution to the entire cosmic positron flux. We note, however, that below 4 
GeV, the MED configuration appears to disagree with the HEAT and AMS 
data, which would favor the MIN configuration.

Thus, we have attempted to discuss the positron fraction 
data by considering two different parameterizations for the electron flux, both 
consistent with the data below 100 GeV. If we had considered other 
fit parameters that were also consistent with the data, by modifying for instance the 
normalizations so as to ensure that they remained correlated 
with the spectral indices\footnote{A correlation 
between the normalization and the spectral index obviously appears in the 
case of a single power law fit, but this would have absolutely no physical 
meaning if there actually were different spectral contributions---secondary, 
far primary, local primary---at different energies to the electron 
flux. A multi-component fit would break this correlation.}, we would have 
found quite different results opening up
more or less the \emph{banana} shape characterizing our predictions
in Fig. \ref{fig:pf-1}. We 
therefore emphasize strongly that it is difficult to interpret the origin of the 
positrons observed in the positron fraction without 
comparing secondary positron predictions with the positron data, or a
precise measurement of the electron flux. This requires 
access to both the electron and positron data separately, which have not 
yet been released in the case of PAMELA. With this in mind, it may be potentially 
unsafe to infer strong statements about the possible nature of any excess, since 
its amplitude and shape depend strongly on the underlying assumptions, which 
are not well constrained at the moment.

To conclude this section, we recall that in the previous sections we have shown 
that our theoretical predictions of the secondary positron flux 
perfectly agree with the available data (PAMELA data for the 
electron and positron components separately are not yet available), especially when the 
theoretical uncertainties are properly taken into consideration. We have 
shown in this section that the electrons play a very important role in the 
interpretation of the positron fraction released by the PAMELA collaboration. 
The scatter in the experimental measurements of the electron flux does not 
allow us, at the moment, to characterize fully the amplitude and shape of any 
positron excess with respect to the secondary yield. We have provided a 
critical illustration of this by using two different parameterizations of the 
electron flux (\emph{hard} and \emph{soft} cases) that are consistent with 
the data below 100 GeV, from which we have shown that the amplitude of an 
excess might be much smaller than expected from the positron/electron model 
of~\cite{1998ApJ...493..694M}. Even if small, this excess could 
originate from additional astrophysical processes or dark matter
annihilation. This implies that it is essential 
to estimate and understand the theoretical uncertainties affecting 
the background, as we have attempted in this paper.

%--------------------------------------------------------------
\section{Conclusions}
\label{sec:conclusion}

Our aim has been to compute a correct estimation of the astrophysical positron flux and the corresponding uncertainties.

First, we have compared the various models available for the interstellar secondary positron production. It has been shown that more positrons are expected when the proton flux from \citet{bess_shikaze_etal_07} is used, as compared to the case proposed by \citet{2001ApJ...563..172D}. Moreover, for a given proton flux, the three positron production cross--sections we have considered produce different results: below a few GeV, the parameterization of~\cite{badhwar_1977} gives more positrons, whereas above a few GeV, the model of~\cite{tan_ng_1983_b} predicts a higher positron production. At any energy, the parameterization of~\cite{Kamae2006} produces the lowest amount of positrons.

Concerning the propagation of the positrons in the interstellar medium, we have used a Green function approach that led us to disregard convection and diffusive reacceleration: we have specifically included diffusion and energy losses due to inverse Compton scattering on cosmic microwave background photons and synchrotron radiation. Nevertheless, this analytical method allowed us to scan our $\sim 1,600$ sets of propagation parameters compatible with the boron to carbon ratio measurements, and therefore to determine astrophysical uncertainties in the positron flux predictions.
We showed that varying the diffusion parameters does not have the same effect as for primary positrons \citep{2008PhRvD..77f3527D}. For exotic positrons created in the Dark Matter halo, the thickness  of the slab $2L$ was the most relevant parameter because the increase of the diffusion zone implies the increase of the number of sources, whereas for secondary positrons -- which are created in the Galactic disk only -- the most relevant parameter is the diffusion constant $K_0$. Therefore, we expect the sets of parameters that basically maximize the primary positron flux to minimize the flux of secondary positrons, and vice versa.

We also showed that, because of energy losses during propagation, most of the positrons detected at the Earth have been created in the nearby 2~kpc: this is the reason why we could safely neglect the variation in proton flux in the Galaxy. By solving the complete equation Eq. (\ref{eq:master_1}) with a numerical technique, we proved that all the other effects (convection, reacceleration, and other losses) can safely be neglected below 10~GeV and that our method is valid.

Finally, and this is our most important result, our estimation of the positron flux is compatible with all available data. This does not mean that there is no exotic positron contribution, since we have not tried to fit the data with a single diffusion model. However, this shows that one should be cautious before claiming that there is any excess in present data. Regarding a possible excess in the positron fraction, we have also clearly shown that the electron flux plays a role that is as important as that of positrons. This might sound tautological because the positron fraction is no more than a ratio, but so much energy is involved in support to this that we repeat this point. Figure~\ref{fig:pf-1} provides compelling visual argument.

The released PAMELA data \citep{PAMELA08} show a clear increase in the positron fraction for energies above 10 GeV. From our analysis, whose objective is an accurate determination of the positron flux, we derive the conclusion that an excess is clear for a hard electron spectrum, while for a soft electron spectrum the rise in the positron fraction may be explained by the standard secondary production. By considering in turn the various parameters, we find in general that the PAMELA measurements are in excess of what a pure secondary component would yield. Nevertheless, if the electron spectrum is soft, most of the PAMELA data points are aligned with our MED prediction. We note also that, in that case, the two last energy bins feature an increase, but the experimental uncertainties are large there and a presence of an excess is, in this case, currently not statisticlly significant.

More insight into these issues will therefore require, from the theoretical side, a revised understanding also of the electron flux, including the determination of its uncertainties, and from the experimental side, the separate provision of the electron and positron fluxes, to allow more robust comparison of theoretical predictions with the data. In addition, the upcoming data on cosmic rays above 10 GeV will allow us to reduce considerably the theoretical uncertainties in all cosmic ray fluxes and help us to elucidate the experimental status of the so-called excess in the positron spectrum.

%--------------------------------------------------------------
%%%%%%%%%%%%%%%%%%%%%%%%%%%%%%%%%%%%%%%%%%%%%%%%%%%%%%%%%%%%%%%%%%%%%%
\begin{acknowledgements}
Work supported by research grants funded jointly by Ministero dell'Istruzione, dell'Universit\`a e della Ricerca (MIUR), by Universit\`a di Torino (UniTO), by
Istituto Nazionale di Fisica Nucleare (INFN) within the {\sl Astroparticle Physics Project}, by the Italian Space Agency (ASI) under contract N$^{\circ}$ I/088/06/0 and by the French Programme National de Cosmologie. T.D. also acknowledges the International Doctorate on AstroParticle Physics (IDAPP) program. R.L. acknowledges financial support from the \emph{Comisi\'on Nacional de Ciencia y Tecnolog\'ia} (CONICYT) from Chile (Grant N$^{\circ}$: BECAS-DOC-BIRF-2005-00) and IDAPP.
\end{acknowledgements}

\bibliographystyle{aa}
\bibliography{bibliography}

%\appendix
\begin{appendix}
\end{appendix}
%%%%%%%%%%%%%%%%%%%%%%%%%%%%%%%%%%%%%%%%%%%%%%%%%%%%%%%%%%%%%%%%%%%%%%

\end{document}